\documentclass[twocolumn]{aastex63}

\newcommand{\mbh}{$M_{\rm BH}$}

\newcommand{\s}{$\sigma$}
\newcommand{\msun}{$M_{\odot}$}
\newcommand{\Lsun}{$L_{\odot}$}
\newcommand{\etal}{et~al.~}

\shorttitle{Black hole mass scaling relations for local active galaxies}
\shortauthors{Bennert et al.}

\begin{document}

\title{A local baseline of the black
  hole mass scaling relations for active galaxies. IV. Correlations
  between \mbh~and host galaxy $\sigma$, stellar mass, and luminosity.}

\correspondingauthor{Vardha N. Bennert}
\email{vbennert@calpoly.edu}

\author{Vardha N. Bennert}
\affiliation{Physics Department, California Polytechnic State University, San
  Luis Obispo, CA 93407, USA}

\author{Tommaso Treu}
\affiliation{Department of Physics, University of California, Los Angeles, CA 90095, USA}

\author{Xuheng Ding}
\affiliation{Kavli Institute for the Physics and Mathematics of the Universe, The University of Tokyo, Kashiwa, Japan 277-8583 (Kavli IPMU, WPI)}
\affiliation{Department of Physics, University of California, Los Angeles, CA 90095, USA}

  \author{Isak Stomberg}
\affiliation{Physics Department, California Polytechnic State University, San
  Luis Obispo, CA 93407, USA}
\affiliation{Department of Physics, KTH,  Royal Institute of Technology, Stockholm, Sweden}
\affiliation{Universit{\"a}t Hamburg, Department of Physics, 20355 Hamburg, Germany}
\affiliation{Deutsches Elektronen-Synchrotron DESY,  22607 Hamburg, Germany}

\author{Simon Birrer}
\affiliation{Kavli Institute for Particle Astrophysics and Cosmology and Department of Physics, Stanford University, Stanford, CA 94305, USA}

\author{Tomas Snyder}
\affiliation{Physics Department, California Polytechnic State University, San
  Luis Obispo, CA 93407, USA}

\author{Matthew A. Malkan}
\affiliation{Department of Physics, University of California, Los Angeles, CA 90095, USA}

  \author{Andrew W. Stephens}
  \affiliation{Gemini Observatory/NSF’s NOIRLab, 670 N. A’ohoku Place, Hilo, Hawai’i, 96720, USA}

\author{Matthew W. Auger}
\affiliation{Institute of Astronomy, Madingley Road, Cambridge CB3 0HA, UK}

\begin{abstract}
  The tight correlations between the mass of supermassive black
holes (\mbh) and their host-galaxy properties have been
of great interest to the astrophysical
community, but a clear understanding of their origin and fundamental
drivers still eludes us. The local relations for active
galaxies are interesting in their own right and form the
foundation for any evolutionary study over cosmic time.  We present Hubble Space Telescope optical imaging  of a sample of 66 local active galactic nuclei
(AGNs); for 14 objects, we also obtained Gemini
  near-infrared images.
We use state of the art methods to perform surface photometry of the
  AGN host galaxies,
  decomposing them in spheroid, disk and bar (when present) and
  inferring the luminosity and stellar mass of the components.
  We combine this information with spatially-resolved kinematics
  obtained at the Keck Telescopes to study the correlations between
\mbh~(determined from single-epoch virial
  estimators) and host galaxy properties.
 The correlations
  are uniformly tight for our AGN sample, with intrinsic scatter
  $0.2-0.4$ dex, smaller than or equal to that of quiescent galaxies.
  We find no difference between pseudo and classical bulges or barred
and non-barred galaxies. We show that all the tight  correlations can be
  simultaneously satisfied by AGN hosts in the 10$^7$-10$^9$
  \msun~regime, with data of sufficient quality.
  The \mbh-$\sigma$ relation is also in agreement with that of
  AGNs with \mbh~obtained from reverberation mapping, providing
 an  indirect validation of single-epoch virial estimators of \mbh.  
\end{abstract}


\section{INTRODUCTION} \label{intro}
When growing through accretion, supermassive black holes (BHs) can be seen as bright nuclei in active galaxies (AGNs).
The observed relations between the mass of the BH (\mbh) and the
properties of the host-galaxy spheroid such as luminosity, stellar mass and
stellar-velocity dispersion $\sigma$,
are thought to result from the co-evolution between BHs and galaxies
\citep[for a review see, e.g.,][]{kor13,gra16}.
Such a co-evolution is either regulated by AGN feedback
\citep[e.g.,][]{dim05,cro06,dub13,dub16,deg15,hop16},
or hierarchical assembly of \mbh~and stellar mass through galaxy
merging \citep[e.g.,][]{pen07,hir10,jah11}.
To shed light on the origin of these relations,
recent years have seen an explosion of observational studies
both in the local Universe
\citep[e.g.,][]{fer05,gre06,gue09,ben11a,kor11,bei12,lae16,dav18,sahu19}
and as a function of cosmic history \citep[e.g.][]{tre04,pen06a,pen06b,woo06,sal07,rie09,jah09,
  ben10,dec10,mer10,ben11b,par15,sex19,sil19,din20}.
  
By necessity, all studies beyond the local Universe
focus on broad-line (or type-1) AGNs (BLAGNs).
For BLAGNs, \mbh~can be estimated to within a factor of 2-3
using empirically calibrated relations based on a sample
of reverberation-mapped AGNs.
Reverberation mapping (RM) is a technique that
studies the time delay between the variability of the accretion disk
and the response of ionized gas in the vicinity of the BH,
the broad-line region (BLR)
\citep[e.g.,][]{wan99,woo02,ves02,ves06,mcg08}.
Using light-travel time arguments,
the time delay translates into a size of the BLR.
Combining the size with the Doppler-broadened width of the emission lines (e.g., the Hydrogen
Balmer series in the optical) results in an estimate of the \mbh~up to  an unknown factor that depends on the geometry and kinematics
of the gas clouds. Traditionally, this factor $f$ (also known as
virial factor) has been derived as a sample average by matching the
scaling relation between \mbh~and (spheroid) stellar-velocity
dispersion \s~of the RM AGNs with that of quiescent galaxies
\citep[e.g.,][]{onk04,par12,woo10,woo15}.
More recently, dynamical modeling of RM data has been used
to constrain both geometry and kinematics of the BLR and thus determine
\mbh~for individual objects, finding consistent results \citep[e.g.,][]{bre11, pan11, li13, pan18,wil18,wil20}.
While RM is time-consuming, the RM AGN sample revealed a relation between BLR size and AGN
luminosity that can be used to estimate \mbh~for BLAGNs
from one spectrum, known as the single-epoch method.
In the single-epoch virial estimation, the width of broad emission lines is combined with the AGN
luminosity which serves as a proxy for BLR size.
As such, the RM AGN sample serves as a \mbh~calibrator beyond the
local Universe.
The single-epoch method has been used for virial mass estimates of
hundreds of thousands of AGNs \citep[e.g.,][]{rak20}, across cosmic history
\citep[e.g.,][]{mor11}, to study the cosmic evolution of the \mbh~scaling
relations \citep[e.g.,][]{tre04, pen06a, woo06, ben10, mer10, par15, din20}
and distribution of Eddington ratios \citep[e.g.,][]{she13}.

Studies of the evolution of the \mbh-host-galaxy scaling relations with redshift constrain theoretical
interpretations and shed light onto their origin
\citep[e.g.,][]{cro06, hop07}; however, they depend on our understanding of the slope and intrinsic scatter of local
relations, in particular those for active galaxies.
Moreover, studying dependencies of the correlations on
bulge structure and other morphological components
at high-redshifts is difficult if not impossible,
especially given the presence of the bright AGN point source in the center.
Late-type galaxies are often known to host pseudo-bulges,
characterized by exponential light profiles, ongoing star
formation or starbursts, and nuclear bars.  It is generally
believed that they have evolved secularly through dissipative
processes rather than mergers
\citep[e.g.,][]{cou96,kor04}.  Classical bulges, in contrast, are thought
of as centrally concentrated, mostly red and quiescent,
merger-induced systems.
Pseudo-bulges and minor mergers 
provide a valuable test of some hypotheses for the origin of the
\mbh~scaling relations:
if they lie on the relation, as found by our results
based on SDSS images \citep{ben15},
it could indicate that secular evolution has a synchronizing effect,
growing BHs and bulges simultaneously
at a small but steady rate for late-type galaxies
\citep{cis11a,cis11b}.

This paper is the last of a series aimed at
creating a robust local baseline of the \mbh~scaling relations of BLAGNs for comparison with 
high redshift studies. We selected
a sample of $\sim$100 Seyfert-1 galaxies from SDSS
(0.02 $\le$ $z$ $\le$ 0.1; \mbh$>10^{7}$M$_{\odot}$) based on their broad H$\beta$ emission 
in the same fashion as high-redshift samples used for evolutionary
studies \citep{ben10,par15,din20}, allowing for a direct comparison.
The majority of AGNs ($\sim$80\%) in our sample reside in galaxies classified as Sa or later
\citep[][]{ben15}, comparable to our high-redshift samples
\citep{ben10,par15}, perhaps not surprisingly,
given that all studies focus on Seyfert-1 galaxies.
In paper I and III \citep[][]{ben11a,ben15}, multi-filter SDSS images yielded photometric
parameters such as the spheroid effective radius, the spheroid
luminosity, the host-galaxy free 5100\AA~luminosity of the AGN
(for an accurate \mbh~measurement), and spheroid stellar masses.
In paper II \citep[][]{har12}, high-quality long-slit Keck/LRIS spectra provided both \mbh~estimates
as well as accurate spatially-resolved stellar-velocity dispersions ($\sigma$)
and rotation curves.

Given the wide range of \mbh, host-galaxy morphologies, and stellar
masses, our sample is well suited to determine the slope and intrinsic scatter of the local scaling relations
and to study dependencies on other parameters such as bulge structure and mergers.
However, relying on low-quality optical photometry (such as
SDSS) whose insufficient angular resolution and limited sensitivity to dust extinction
significantly increase observational scatter in the \mbh~scaling relations,
ultimately limits conclusive results.
High-resolution images are essential
to resolve (pseudo-) bulges, given that roughly half of all objects
have bulge effective radii smaller than $\sim$1.5$\arcsec$
\citep[corresponding to $\sim$1.7 kpc for a typical distance of the
galaxies in our sample;][]{ben15}.
This is not only crucial for proper morphological classification
and determination of bulge luminosity,
it also is important for the effective radius measurement.
The latter, in turn, is important for a robust measurement of
spatially-resolved stellar-velocity dispersion within the effective
spheroid radius \citep{ben15}.
In other words, angular resolution is the key for
an accurate determination of all \mbh~scaling relations.

In this paper,  to overcome these problems and to obtain high-quality host-galaxy images,
we took a two-pronged approach.
A sub-set of the parent sample (15 objects), selected
to cover a wide range of morphologies (as based on SDSS images),
was observed with the Near InfraRed Imager and spectrograph (NIRI) on
Gemini North. Gemini-NIRI was chosen
(i) for its high-spatial resolution
(instrument plus site seeing)
to distinguish between classical and pseudo-bulges
in the presence of an AGN point source;
(ii) for its large field-of-view
($2\arcmin \times 2\arcmin$ at f/6) 
to measure the surface brightness profile of
these nearby galaxies out to large radii;
and (iii) because near-infrared observations maximize
 the contrast between AGN and host and minimize dust extinction, 
 revealing the presence of (pseudo-) bulges, bars and (minor) mergers.
The reduced dust extinction also makes NIR luminosities a better tracer of stellar mass.
At the same time, the parent sample was part of a Hubble Space
Telescope (HST) snapshot (SNAP) program
(PI Bennert). HST images provide both a high spatial resolution and a stable
point-spread function (PSF). 
WFC3/UVIS was used with broad-band filter F814W
to maximize the contrast between AGN and host, avoiding strong AGN
emission lines, while taking full advantage of the high resolution of UVIS.
Compared to existing SDSS images, the
HST images have a factor of $\sim$40 increase in resolution.
A total of 66 objects were observed with HST,
14 of which also have Gemini images.
Gemini and HST images naturally complement each other and
together provide a long wavelength range for stellar-mass determination for
overlapping objects. By construction, SDSS images in five filters are available for
all objects to further assist in constraining stellar masses.
Combining the high-quality spectroscopic data (paper II) with high-quality
imaging provides a representative
spectral and spatial coverage of supermassive BHs and their hosts for a detailed
mapping of the local \mbh~scaling relations for active galaxies and
their underlying drivers.

The paper is organized as follows.
Section~\ref{sample} summarizes the sample selection,
HST and Gemini observations, and
data reduction.  Section~\ref{analysis} describes the
analysis and derived quantities. Section~\ref{results} presents
host-galaxy morphologies and discusses the resulting \mbh~scaling relations.
Section~\ref{summary} concludes with a summary.
Throughout this paper, magnitudes are given in AB magnitudes. For conversion
to luminosities, absolute solar magnitudes were taken from
\citet{will18} and a Hubble constant of H$_{\rm{0}}$ = 70 km s$^{-1}$ and a flat Universe
with a cosmological constant of $\Omega_{\rm{\lambda}}$ = 0.7 are assumed.

\section{SAMPLE SELECTION, OBSERVATIONS, AND DATA REDUCTION}
\label{sample}
\subsection{Sample selection}
The parent sample is 102 type-1 Seyfert
galaxies selected from the Sloan
Digital Sky Survey (SDSS) data release six 
\citep{ade08} based on
redshift (0.02 $\leq z \leq$ 0.1) and \mbh~($>10^{7}M_{\odot}$), and
observed with Keck/LRIS, presented in detail in papers I, II and III in this series
\citep{ben11a,har12,ben15}.
Accurate spatially-resolved stellar-velocity dispersions were obtained
for 84 objects (paper II) and formed the sample for our HST snapshot (SNAP) program.
15 objects with a wide variety of host-galaxy properties, as
determined from SDSS images \citep{ben15}, were observed with NIRI on
Gemini North (PI: Bennert; program ID GN-2016B-Q-33).
68 objects were observed as part of HST SNAP, although we were unable
to determine a robust \mbh~for 2 of them due to a lack of broad
H$\beta$ in the Keck spectrum \citep[despite it being present in prior
SDSS spectra;][]{run16}.
Of the remaining 66, 14 overlap with the Gemini sample.
Fully-reduced SDSS images are available for all objects
through the SDSS archive.
Sample properties (coordinates, redshift and host-galaxy
  morphology) can be found
in Table~\ref{table:sample}.

\startlongtable
\begin{deluxetable*}{lccccccccc}
\label{table:sample}
\tabletypesize{\footnotesize}
\tablecolumns{10}
\tablecaption{Sample and host-galaxy properties}
\tablehead{\colhead{Object} & 
\colhead{R.A.} &
\colhead{Decl.} &
\colhead{$z$} &
\colhead{Frame} &
\colhead{Host}
& \colhead{$n$}
& \colhead{B2T}
& \colhead{rot.}
& \colhead{bar}
                            \\
& (J2000) & (J2000) & & ('')   \\
(1) & (2)  & (3) & (4) & (5) & (6) & (7) & (8) & (9) & (10) 
                                                                     }
                                                                   \startdata
0013-0951 & 00 13 35.38 & -09 51 20.9 & 0.0615 & 21.7 & BD (C) & Y & Y & N & N \\
0038+0034 & 00 38 47.96 & +00 34 57.5 & 0.0805 & 18.9 & BD (C) & N & N & N & N \\
0109+0059 & 01 09 39.01 & +00 59 50.4 & 0.0928 & 14.0 & BDB (P) & Y & Y & N & Y \\
0121-0102 & 01 21 59.81 & -01 02 24.4 & 0.054 & 28.0 & BDB (P) & Y & Y & N & Y \\
0150+0057 & 01 50 16.43 & +00 57 01.9 & 0.0847 & 21.7 & BDB (P) & Y & Y & N & Y \\
0206-0017 & 02 06 15.98 & -00 17 29.1 & 0.043 & 56.0 & BD (C) & N & N & N & N \\
0212+1406 & 02 12 57.59 & +14 06 10.0 & 0.0618 & 21.7 & BDB (P) & Y & Y & N & Y \\
0301+0110 & 03 01 24.20 & +01 10 22.1 & 0.0715 & 14.0 & BD (C) & N & N & N & Y \\
0301+0115 & 03 01 44.19 & +01 15 30.8 & 0.0747 & 12.6 & BDB (P) & Y & Y & N & Y \\
0336-0706 & 03 36 02.09 & -07 06 17.1 & 0.097 & 23.8 & BD (P) & Y & Y & N & Y \\
0353-0623 & 03 53 01.02 & -06 23 26.3 & 0.076 & 19.6 & BDB (C) & Y & Y & N & N \\
0737+4244 & 07 37 03.28 & +42 44 14.6 & 0.0882 & 14.0 & BD (C) & N & Y & N & N \\
0802+3104 & 08 02 43.40 & +31 04 03.3 & 0.0409 & 19.6 & BDB (P) & Y & Y & N & Y \\
0811+1739 & 08 11 10.28 & +17 39 43.9 & 0.0649 & 21.7 & BDB (P) & Y & Y & N & Y \\
0813+4608 & 08 13 19.34 & +46 08 49.5 & 0.054 & 23.8 & BDB (C) & N & Y & N & Y \\
0845+3409 & 08 45 56.67 & +34 09 36.3 & 0.0655 & 28.0 & BDB (C) & N & Y & N & Y \\
0857+0528 & 08 57 37.77 & +05 28 21.3 & 0.0586 & 19.6 & BD (C) & Y & Y & N & N \\
0904+5536 & 09 04 36.95 & +55 36 02.5 & 0.0371 & 31.5 & BD (C) & Y & N & N & N \\
0909+1330 & 09 09 02.37 & +13 30 18.2 & 0.0506 & 31.5 & BDB (P) & Y & Y & N & Y \\
0921+1017 & 09 21 15.55 & +10 17 40.9 & 0.0392 & 39.2 & BD (C) & N & Y & N & N \\
0923+2254 & 09 23 43.00 & +22 54 32.7 & 0.0332 & 47.6 & BDB (P) & Y & Y & N & Y \\
0923+2946 & 09 23 19.73 & +29 46 09.1 & 0.0625 & 19.6 & B (C) & N & N & N & N \\
0927+2301 & 09 27 18.51 & +23 01 12.3 & 0.0262 & 59.5 & BD (C) & Y & Y & N & N \\
0932+0233 & 09 32 40.55 & +02 33 32.6 & 0.0567 & 17.5 & BD (C) & Y & Y & N & N \\
0936+1014 & 09 36 41.08 & +10 14 15.7 & 0.06 & 39.2 & BD (C) & Y & Y & N & N \\
1029+1408 & 10 29 25.73 & +14 08 23.2 & 0.0608 & 28.0 & BD (C) & N & N & N & N \\
1029+2728 & 10 29 01.63 & +27 28 51.2 & 0.0377 & 19.6 & BD (C) & N & N & N & N \\
1029+4019 & 10 29 46.80 & +40 19 13.8 & 0.0672 & 15.4 & BD (C) & Y & Y & N & N \\
1042+0414 & 10 42 52.94 & +04 14 41.1 & 0.0524 & 17.5 & BDB (P) & Y & Y & N & Y \\
1043+1105 & 10 43 26.47 & +11 05 24.3 & 0.0475 & 14.0 & B (C) & N & N & N & N \\
1058+5259 & 10 58 28.76 & +52 59 29.0 & 0.0676 & 23.8 & BDB (P) & Y & Y & N & Y \\
1101+1102 & 11 01 01.78 & +11 02 48.8 & 0.0355 & 28.0 & BD (C) & N & N & Y & N \\
1104+4334 & 11 04 56.03 & +43 34 09.1 & 0.0493 & 12.6 & BDB (C) & N & N & N & Y \\
1137+4826 & 11 37 04.17 & +48 26 59.2 & 0.0541 & 7.7 & BD (C) & N & N & N & N \\
1143+5941 & 11 43 44.30 & +59 41 12.4 & 0.0629 & 31.5 & BDB (C) & N & Y & N & Y \\
1144+3653 & 11 44 29.88 & +36 53 08.5 & 0.038 & 31.5 & BD (C) & Y & Y & N & N \\
1145+5547 & 11 45 45.18 & +55 47 59.6 & 0.0534 & 28.0 & BDB (C) & Y & Y & N & N \\
1147+0902 & 11 47 55.08 & +09 02 28.8 & 0.0688 & 19.6 & BD (C) & N & N & N & N \\
1205+4959 & 12 05 56.01 & +49 59 56.4 & 0.063 & 28.0 & BD (C) & N & Y & N & N \\
1206+4244 & 12 06 26.20 & +42 44 26.95 & 0.052 & 28.0 & BDB (P) & Y & Y & N & Y \\
1216+5049 & 12 16 07.09 & +50 49 30.0 & 0.0308 & 47.6 & BD (C) & N & Y & N & N \\
1223+0240 & 12 23 24.14 & +02 40 44.4 & 0.0235 & 28.0 & BD (C) & N & Y & N & N \\
1246+5134 & 12 46 38.74 & +51 34 55.9 & 0.0668 & 17.5 & BD (P) & Y & Y & Y & N \\
1306+4552 & 13 06 19.83 & +45 52 24.2 & 0.0507 & 23.8 & BDB (P) & Y & Y & N & Y \\
1307+0952 & 13 07 21.93 & +09 52 09.3 & 0.049 & 23.8 & BDB (P) & Y & Y & N & Y \\
1312+2628 & 13 12 59.59 & +26 28 24.0 & 0.0604 & 28.0 & BDB (P) & Y & Y & N & Y \\
1405-0259 & 14 05 14.86 & -02 59 01.2 & 0.0541 & 23.8 & BD (C) & Y & Y & N & N \\
1416+0137 & 14 16 30.82 & +01 37 07.9 & 0.0538 & 39.2 & BD (C) & N & Y & N & N \\
1419+0754 & 14 19 08.30 & +07 54 49.6 & 0.0558 & 35.7 & BD (P) & Y & Y & Y & N \\
1434+4839 & 14 34 52.45 & +48 39 42.8 & 0.0365 & 31.5 & BDB (P) & Y & Y & N & Y \\
1545+1709 & 15 45 07.53 & +17 09 51.1 & 0.0481 & 15.4 & BD (C) & N & N & N & N \\
1557+0830 & 15 57 33.13 & +08 30 42.9 & 0.0465 & 7.7 & B (C) & N & N & N & N \\
1605+3305 & 16 05 02.46 & +33 05 44.8 & 0.0532 & 15.4 & BD (C) & Y & Y & N & N \\
1606+3324 & 16 06 55.94 & +33 24 00.3 & 0.0585 & 26.6 & BD (C) & N & N & N & N \\
1611+5211 & 16 11 56.30 & +52 11 16.8 & 0.0409 & 19.6 & BD (C) & N & N & N & N \\
1636+4202 & 16 36 31.28 & +42 02 42.5 & 0.061 & 29.4 & BDB (C) & N & N & N & Y \\
1708+2153 & 17 08 59.15 & +21 53 08.1 & 0.0722 & 28.0 & BD (P) & Y & Y & Y & N \\
2116+1102 & 21 16 46.33 & +11 02 37.3 & 0.0805 & 23.8 & BDB (P) & Y & Y & N & Y \\
2140+0025 & 21 40 54.55 & +00 25 38.2 & 0.0838 & 14.0 & BD (C) & Y & Y & N & N \\
2215-0036 & 22 15 42.29 & -00 36 09.6 & 0.0992 & 19.6 & BD (C) & Y & Y & N & N \\
2221-0906 & 22 21 10.83 & -09 06 22.0 & 0.0912 & 12.6 & BD (C) & N & Y & N & N \\
2222-0819 & 22 22 46.61 & -08 19 43.9 & 0.0821 & 19.6 & BDB (P) & Y & Y & N & Y \\
2233+1312 & 22 33 38.42 & +13 12 43.5 & 0.0934 & 31.5 & BDB (P) & Y & Y & N & Y \\
2254+0046 & 22 54 52.24 & +00 46 31.4 & 0.0907 & 31.5 & BD (C) & Y & N & N & N \\
2327+1524 & 23 27 21.97 & +15 24 37.4 & 0.0458 & 31.5 & BD (C) & N & N & N & N \\
2351+1552 & 23 51 28.75 & +15 52 59.1 & 0.0963 & 19.6 & BD (C) & N & Y & N & N \\
  \enddata
  \tablecomments{
Col. (1): Target ID used throughout the text (based on R.A. and declination). 		      	   	  
Col. (2): Right ascension in hours, minutes and seconds.				  
Col. (3): Declination in degrees, arcminutes and arcseconds.
Col. (4): Redshift from SDSS data release seven \citep{aba09}.
Col. (5): Frame size of image shown in Figs.~\ref{figure:hst1}-\ref{figure:hst3}  (in
arcsecond, in x and y).
Col. (6): Host-galaxy fit (B: spheroid only, BD: spheroid+disk, BDB:
spheroid+disk+bar). In parentheses: Spheroid component: C = classical
bulge; P = pseudo-bulge.
Col. (7-10): Criteria for classification of bulge as pseudo-bulge. If
at least three of these four criteria are met, the bulge is classified as pseudo-bulge.
Col. (7): S{\'e}rsic index $n$ $<$ 2.
Col. (8): Bulge-to-total luminosity ratio $<$ 0.5.
Col. (9): Rotation dominated (i.e., ratio between maximum rotational
velocity at effective spheroid radius and central stellar-velocity
dispersion $>$ 1).
Col. (10): Presence of a bar for face-on galaxies.}
\end{deluxetable*}

\subsection{HST observations and data reduction}
84 objects were part of an HST SNAP program 
(HST GO 15215 and HST GO 16014, PI Bennert; Cycles
25-27), yielding images for 68 AGN host galaxies, a completion rate of
almost 80\% (significantly higher than the 30\% typical for SNAP programs).
To obtain the dynamic range needed for an
accurate decomposition of the host galaxy and the AGN, the long
exposures (400 seconds) were complemented by short, unsaturated ones
(between 20-100 seconds, depending on the object's brightness).
To avoid buffer dump (which occurred during the long exposures),
a sequence of one short and one long exposure at the same
location was followed by another sequence of one long and one short
exposure at a dithered location.
POS TARG was used to set up a dither pattern manually that corresponds to the
default WFC3-UVIS-GAP-LINE (with center UVIS).
Full-frame images were read  to trace the host-galaxy disks out
to the background and
to obtain field stars for PSF fitting of the strong AGN point source
in the center.

Data processed through the standard WFC3 calibration pipeline were
retrieved from the HST archive.
L.A. Cosmic (Laplacian Cosmic Ray Identification) \citep{vanDok01}
was run to remove cosmic rays. All long exposures were carefully
checked for saturation, especially of the bright AGN point source. For
objects with saturated pixels, the short exposures taken at the same
dither location as the long ones were scaled according to exposure
time and used to replace the saturated pixels.  Pyraf package
astrodrizzle was then used to combine the two long exposures.
A wide range of combinations of the final drizzle parameters scale and pixfrac
were applied. After careful examination of the images and based on
resolution, image quality and FWHM of the PSF, the following parameters were adopted:
driz$\_$sep$\_$bits=336, final$\_$bits=336, final$\_$wcs=yes, final$\_$pixfrac=0.9,
final$\_$scale=0.035,
resulting in a final pixel scale of 0.035$\arcsec$/pix.
For objects without any saturated pixels, the long exposures were combined with astrodrizzle directly in the same way.
Final images are shown in Figures~\ref{figure:hst1}-\ref{figure:hst3}.

\begin{figure*}
  \center
    \includegraphics[scale=0.33]{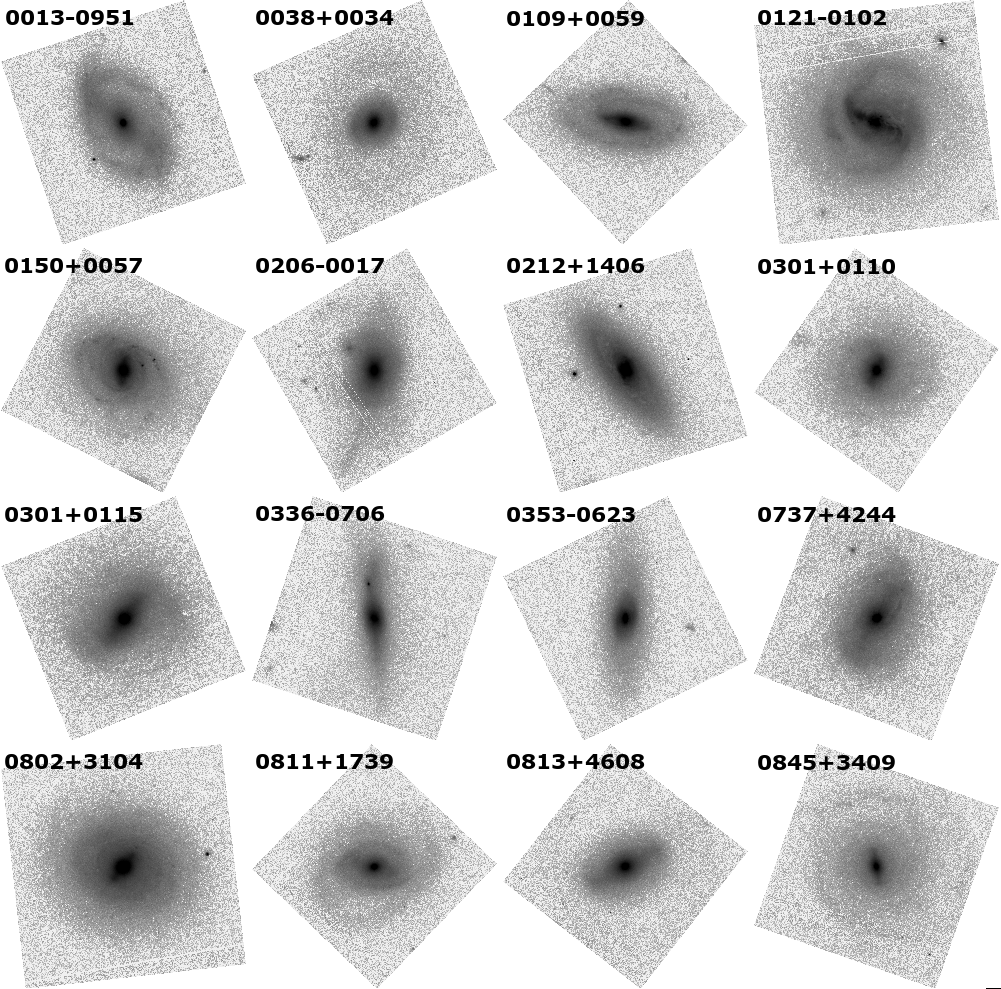}\\
    \includegraphics[scale=0.33]{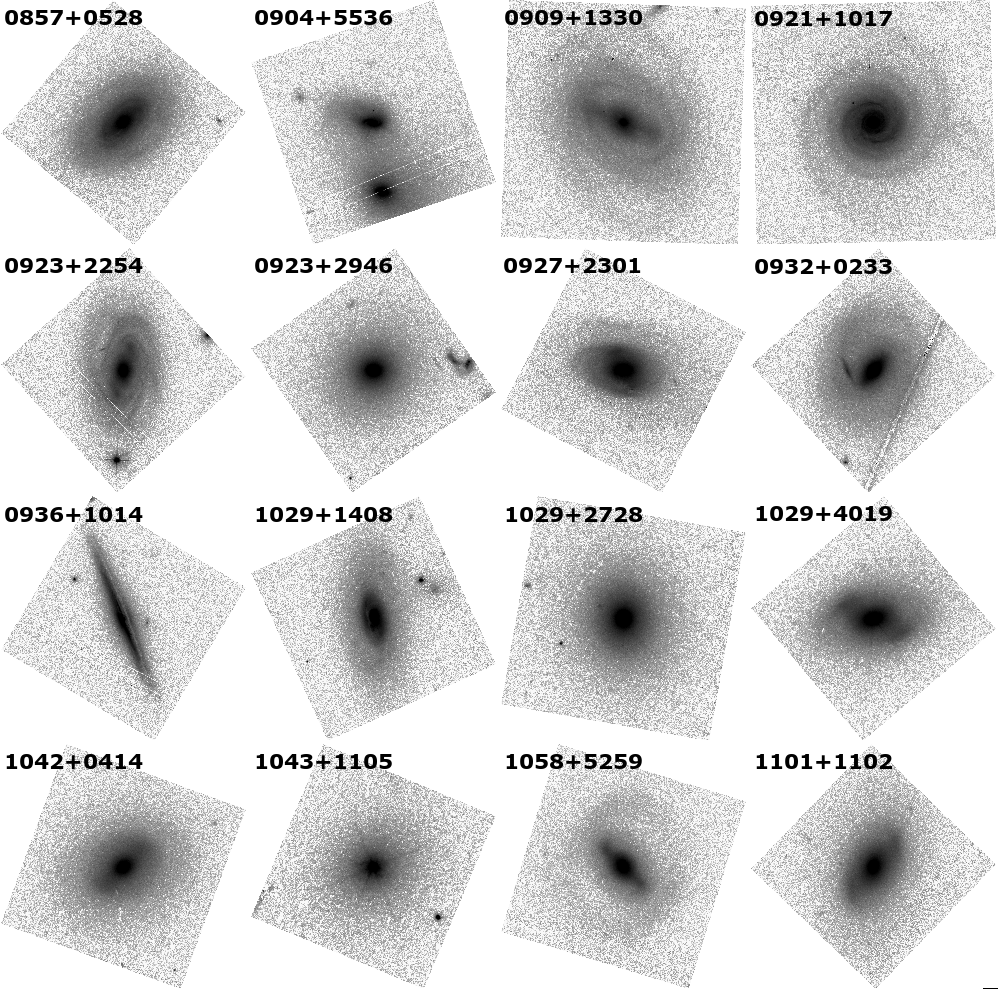}
    \caption{HST UVIS/F814W images of our sample.
      Cutouts are shown as used for the fitting, but rotated here
(North up and East to the left) for display purposes.
      Image sizes
      are listed in
      table~\ref{table:sample}. Thanks to the high spatial resolution
      and S/N data, the wide variety of host-galaxy
      morphologies can clearly be seen.}
  \label{figure:hst1}
  \end{figure*}
  
  \begin{figure*}
    \center
    \includegraphics[scale=0.33]{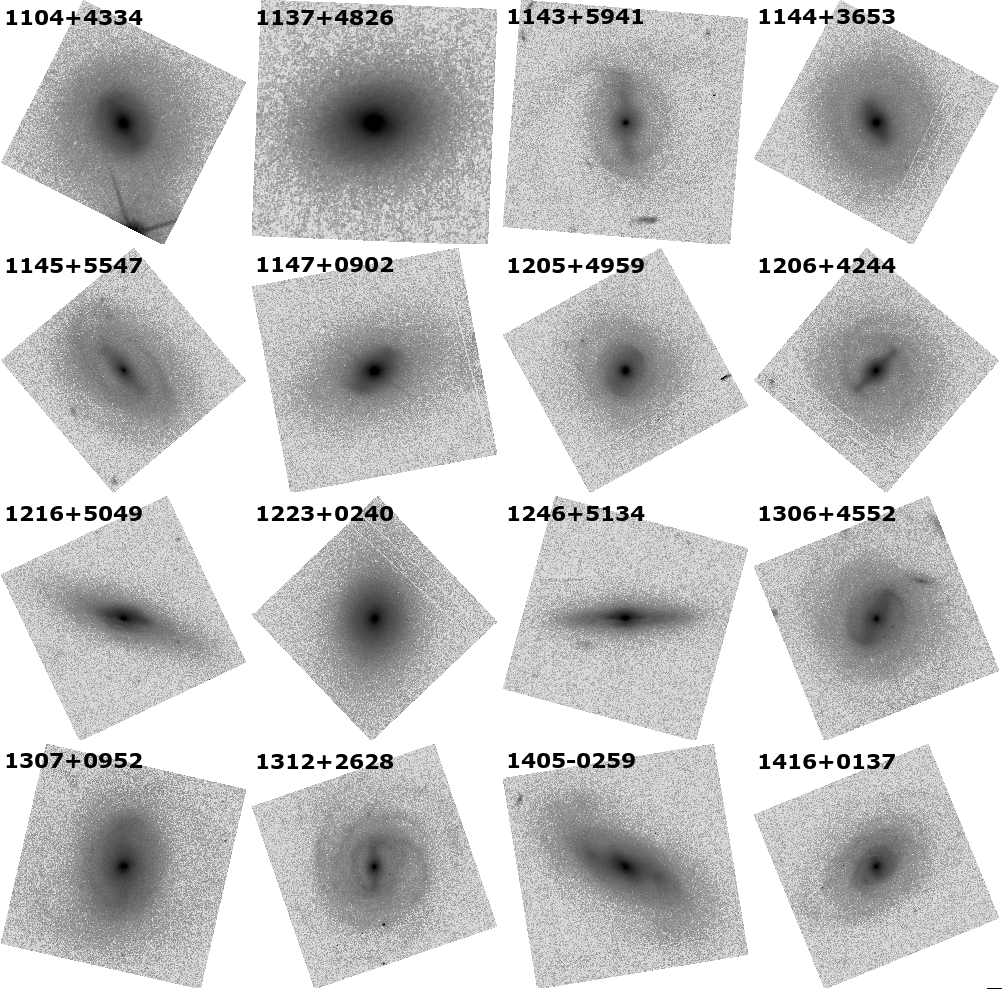}\\
    \includegraphics[scale=0.33]{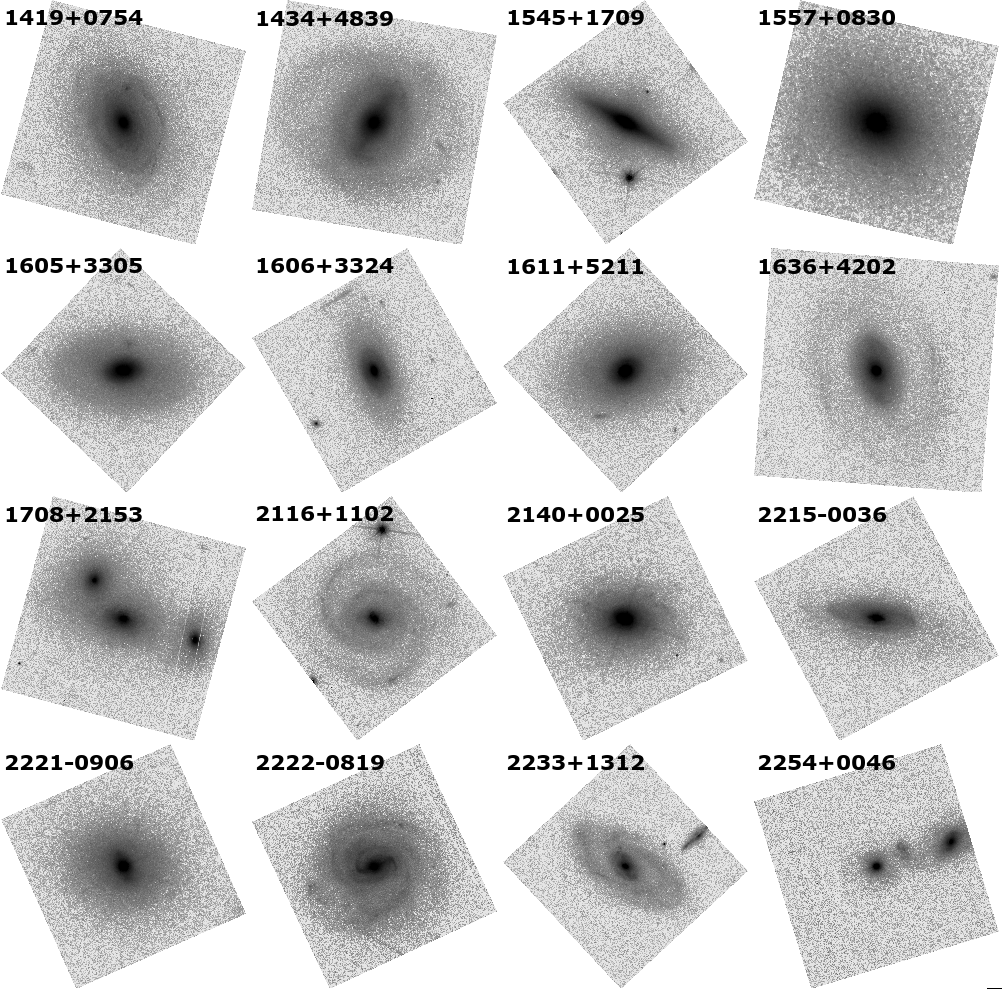}
   \caption{Continuation of Figure~\ref{figure:hst1}.}
    \label{figure:hst2}
  \end{figure*}

    \begin{figure*}
    \center
    \includegraphics[scale=0.33]{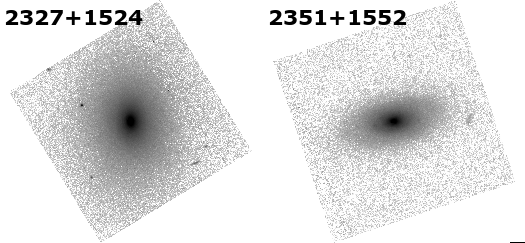}
   \caption{Continuation of Figure~\ref{figure:hst1}.}
    \label{figure:hst3}
  \end{figure*}

\subsection{Gemini observations and data reduction}
15 Seyfert-1 galaxies were selected from the parent sample
covering a wide range of morphologies (based on SDSS images).
All galaxies were observed with NIRI on Gemini North
with the largest field-of-view (FOV; $2\arcmin \times 2\arcmin$
at f/6; pixel scale of 0.117$\arcsec$) in Ks band.
Observations have an average FWHM of 0.33$\arcsec$
(0.24 - 0.44$\arcsec$)
and were obtained at airmass less than 1.5.  This image quality is 3-4
times better than the SDSS images which have a typical seeing of
1.5$\arcsec$.
Exposure times range between 2-10 seconds per image with 3-6 co-adds
and 18-24 images on source, resulting in a total exposure time
of 144-540 seconds, depending on the brightness of the object.
The sky positions were observed with guiding enabled and were
carefully selected to include a nearby bright field star.
This ensured that we could generate accurate PSF models for every galaxy.
Details of the observations are given in Table~\ref{table:gemini}.

\begin{figure*}
  \center
      \includegraphics[scale=0.33]{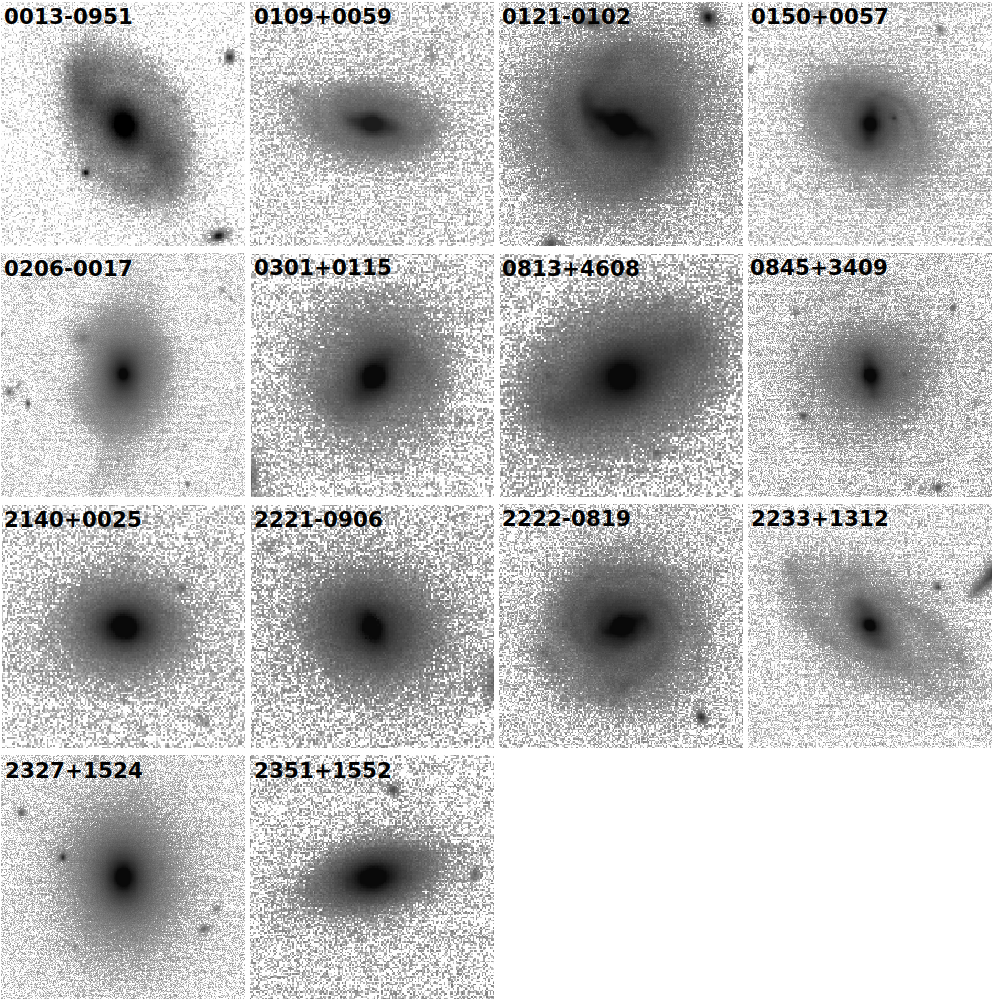}
    \caption{Gemini NIRI Ks images for 14 galaxies. North is up, East
      is to the left. Image sizes are listed in
      table~\ref{table:gemini}.}
    \label{figure:gemini}
    \end{figure*}

Data reduction was performed following standard procedures
using the Gemini IRAF \citep{tod86,tod93} package customized for NIRI
and included dark subtraction and flat fielding using
off-target exposures.
Flux calibration was obtained
by standard IRAF photometry of
UKIRT faint standard stars observed
directly before or after the science images.
Absolute magnitudes take into account extinction
\citep{sch11};
Ks-band luminosities were determined assuming an absolute Ks-band magnitude of the Sun
of $M_{\rm Ks}$ = 5.08 \citep{will18}
Resulting images are shown in Figure~\ref{figure:gemini}.
  
\begin{deluxetable*}{lcccc}
\label{table:gemini}
\tabletypesize{\footnotesize}
\tablecolumns{15}
\tablecaption{Gemini Observations.}
\tablehead{
\colhead{Object}  & 
          \colhead{Date of Obs.}  &
                   \colhead{Exp. time}  & \colhead{FWHM}  &
                   \colhead{Frame size} \\
& (UT)  & (s)  & ('') & ('') \\
(1)  & (2)   & (3)  & (4)  & (5)}
\startdata
0013$-$0951 & 2016 Oct 25  & 378 & 0.31 & 21.1\\ 
0109+0059   & 2016 Oct 14  & 360 & 0.32 & 18.7\\ 
0121$-$0102 & 2016 Sep 06  & 360 & 0.35 & 23.4\\ 
0150+0057   & 2017 Jan 09  & 216 & 0.27 & 23.4\\ 
0206$-$0017 & 2017 Jan 09  & 144 & 0.24 & 46.8\\ 
0301+0115   & 2016 Nov 06  & 360 & 0.36 & 16.4\\ 
0813+4608   & 2016 Oct 20  & 360 & 0.30 & 15.2\\ 
0845+3409   & 2016 Oct 30  & 324 & 0.40 & 28.1\\ 
2140+0025   & 2016 July 16 & 360 & 0.29 & 16.4\\ 
2221$-$0906 & 2016 July 16 & 540 & 0.32 & 16.4\\ 
2222$-$0819 & 2016 Sep 08  & 378 & 0.27 & 21.1\\ 
2233+1312   & 2016 Aug 7   & 216 & 0.32 & 25.7\\ 
2327+1524   & 2016 Aug 8   & 216 & 0.44 & 35.1\\ 
  2351+1552   & 2016 Oct 15  & 432 & 0.26 & 18.7\\
  \enddata
  \tablecomments{
  Col. (1): Target ID used throughout the text (based on R.A. and declination).
Col. (2): Date of observation (UT).
Col. (3): Total exposure time in seconds.
Col. (4): Full-Width at Half Maximum (FWHM) of PSF star in arcsecond.
Col. (5): Frame size of image shown in Fig.~\ref{figure:gemini} (in
arcsecond, in x and y).}
\end{deluxetable*}

\section{DERIVED QUANTITIES}
\label{analysis}

\subsection{Surface photometry}
\label{surface}
To perform a detailed 2D host-galaxy fitting, we use the public image analysis software 
lenstronomy\footnote{\url{https://github.com/sibirrer/lenstronomy}}
\citep{bir18}. Lenstronomy supersedes GALFIT \citep{pen02} by applying an
MCMC technique to provide realistic errors and explore the covariance
between the various model parameters. It allows for a more general
surface brightness reconstruction possible with a large non-parametric
basis set; the coefficients are determined through a linear
minimization rather than a non-linear parameter fitting
\citep{bir15}. Also, iterative PSF reconstruction is possible and
allows one to incorporate residual uncertainties due to PSF mismatch
into the analysis. While lenstronomy was originally developed for
galaxy-scale strong gravitational lensing, it has a much broader
application, including general 2D galaxy decompositions.

A well subtracted background and a matching PSF is important
for obtaining reliable host galaxy properties from 2D
surface photometry.
We estimate and remove the local background light in 2D space based on
the SExtractorBackground
algorithm built in the photutils package (Python based), which effectively accounts for gradients in the background light distribution.
For all objects, PSF stars were created from suitable stars in the
field-of-view (FOV) of each object, following the criteria: bright,
unsaturated star without any nearby objects, located close to the
AGN/center of the FOV and an overall profile as expected for a PSF
star. These individual PSF stars were then recentered and stacked,
resulting in a PSF with high signal-to-noise ratio (S/N), centered in the image. Each individual PSF star was 
recentered using
AstroObjectAnalyzer\footnote{\url{https://github.com/sibirrer/AstroObjectAnalyser}}
through 
an iterative interpolation algorithm.
Finally, masks were created to mask any nearby sources. Three objects have close-by neighboring galaxies which were
fitted simultaneously.

The central AGN was fitted by a PSF, 
the host galaxies with three different models:
(1) a spheroid-only component \citep[free S{\'e}rsic index $n$;][]{ser63};
(2) a spheroid plus disk component (S{\'e}rsic index = 1)
(3) a spheroid plus disk plus bar component (S{\'e}rsic index = 0.5).
Based on pre-defined starting parameters and constraints, lenstronomy
determines the maximum likelihood fit adopting
a Particle Swarm Optimizer (PSO) \citep{ken01}.
We use the following limits:
effective radius $r_{\rm eff}$ for all components: 3 $\times$ pixel size $\le r_{\rm eff}$
$\le$ 30 $\times$ pixel size; spheroid S{\'e}rsic index $n$: $1 \le n \le
5$.
Also, for the spheroid-disk fit, we force the disk to be larger than
the spheroid and more elliptical. Likewise, for the spheroid-disk-bar
fit, we force the disk to be larger than the bar and the spheroid,
and the spheroid component to be the most round one of three components.

After running various trials, the two main challenges that we encountered were
(i) determining the best values for the PSO chains, to make sure the
code converged and
(ii) making sure that the
code converged to the true global minimum and not to a local one.
The latter may
depend on the starting parameters used, especially if more than one component
is fitted to the host galaxy, due to degeneracies involved and the high-dimensional parameter volume.
After some experimenting, the following procedure was shown to be
successful.
We chose a PSO chain that guaranteed convergence even for the largest image size.
For a spheroid-only fit, we chose a PSO with 200 particles and 70 iterations,
for spheroid-disk and spheroid-disk-bar, a PSO with 300 particles and
100 iterations. For all fits, as a diagnostic, we displayed the log (likelihood) of the fit,
 particle position and parameter velocity for the different parameters as a function of
iteration to ensure that the chain indeed converged.

\begin{figure*}
   \center
    \includegraphics[scale=0.4]{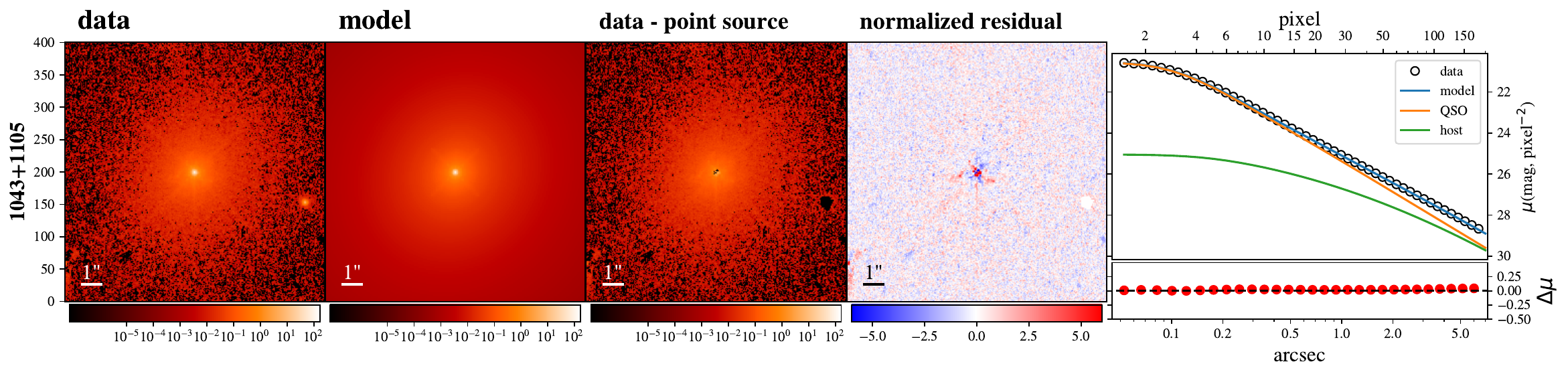}
    \includegraphics[scale=0.4]{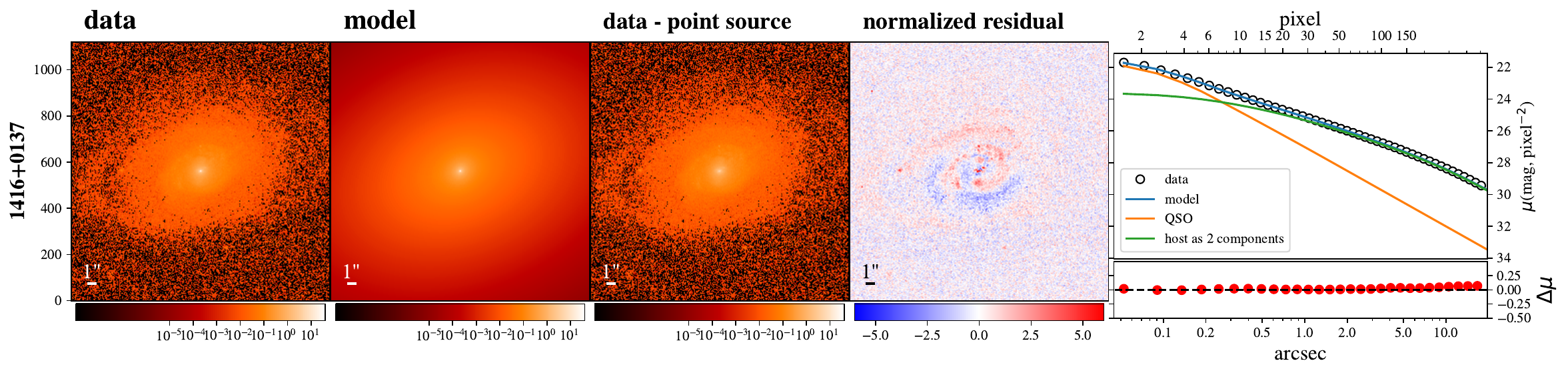}
    \includegraphics[scale=0.4]{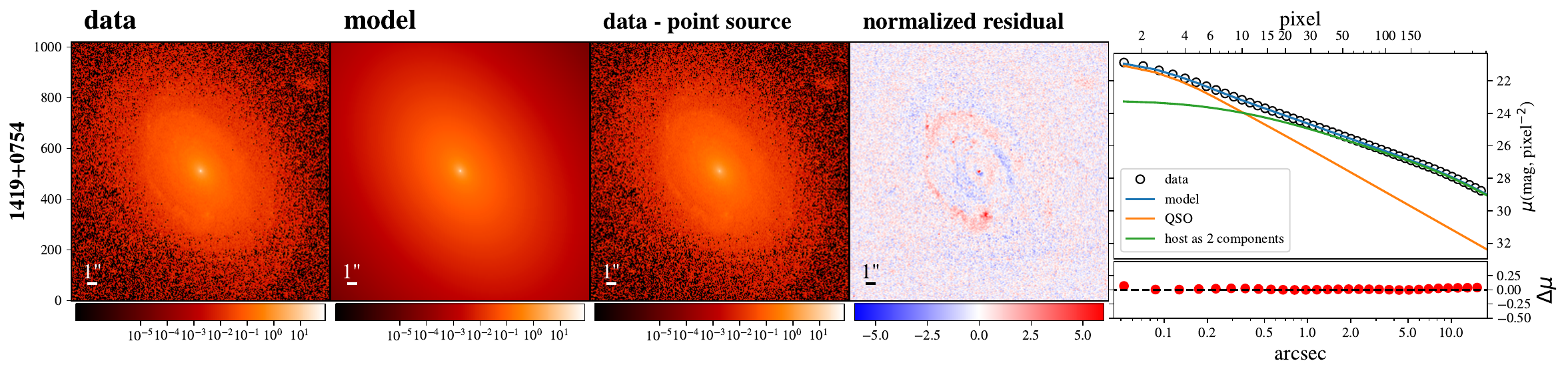}
    \includegraphics[scale=0.4]{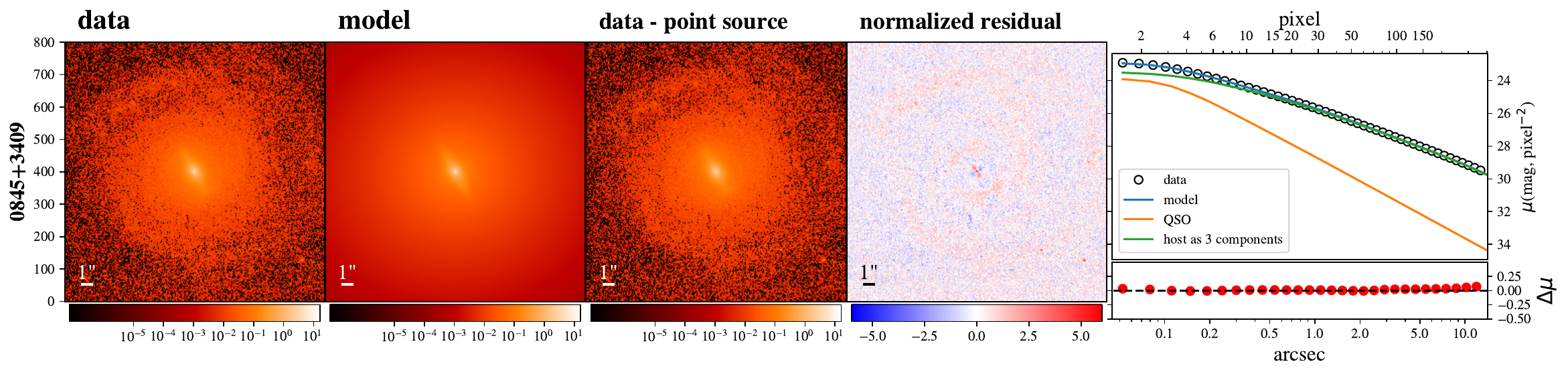}
    \includegraphics[scale=0.4]{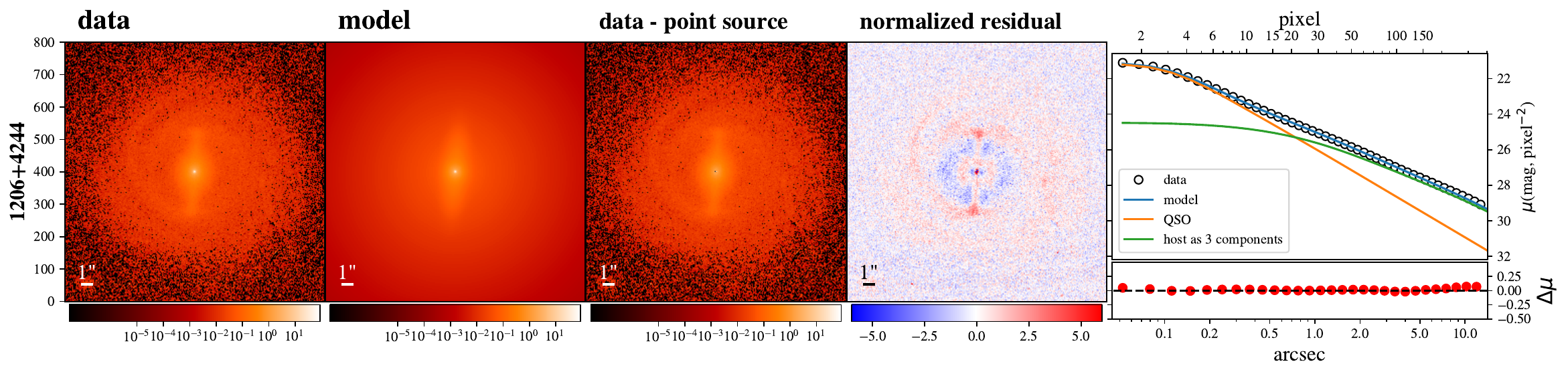}
    \caption{Example fits from lenstronomy.
From left to right: HST image (``data''), best fit model derived from
lenstronomy (``model''); PSF-subtracted image (``data - Point
Source''); residual image after subtraction of best fit model from
data, divided by the noise level (``normalized residual'');
and surface-brightness profile showing the data as black circles,
the model as a blue line, the PSF fitting the central AGN as an orange
line and the host-galaxy total fit as a green line.
Note that the images are shown as observed with HST and as
  fitted by lenstronomy, and as such have random orientation (not rotated with North up and East to the left).
From top to bottom,
five different objects are shown, representing a variety of host-galaxy morphologies:
1043+1105 is an elliptical galaxy fit with a single S{\'e}rsic;
1416+0137 is a disk galaxy with classical bulge;
1419+0754 is a disk galaxy with pseudo-bulge;
0845+3409 is a barred disk galaxy with classical bulge;
and 1206+4244 is a barred disk galaxy with pseudo-bulge.
(Note that the surface-brightness plots in the right panels are shown only for
illustrative purposes; the fit was performed on the 2D
image. Surface-brightness values are given in the plane of the sky;
the x-axis is based on a circularized radius.)
}
\label{lenstronomyfits}
\end{figure*}

First, we ran the spheroid-only fit which was shown to be robust,
i.e.~giving the same fitting result, regardless of starting
parameters used.
We then used the results from the spheroid-only fit as starting
parameters for the disk in the subsequent spheroid-disk and
spheroid-disk-bar fit.
For objects for which there is at least an indication of a visual bar,
the bar parameters size, position angle and ellipticity were
carefully  determined manually and used as starting parameters for the
bar in the spheroid-disk-bar fit.
For both the spheroid-disk and spheroid-disk-bar fit, we
chose three different starting parameters for the size of the spheroid
and the disk, to cover a broader range.
For the spheroid effective radius, we chose three different starting parameters:
(i) pixel size * 4, (ii) pixel size * 8, and (iii) pixel size *30.
For the disk effective radius we chose these three different starting
parameters:
(i) spheroid effective radius
from spheroid-only fit, (ii) twice the size used in (i), and (iii) half
the size used in (i).
When combined, this yields nine different starting parameters for
both the spheroid-disk and spheroid-disk-bar fits.
The nine fits were compared in terms of image residuals and resulting
chi-squared values. In most cases, the nine different fits yielded 
identical results, showing the convergence to a true global minimum.
Occasionally, outliers were identified through higher chi-square values
and/or from the residual image and excluded.
By careful inspection of the images and final fitting results, 
we determined the best model and fit for all objects.
A disk and bar were included in the host-galaxy fitting only if they
were clearly visible in the image and/or if their inclusion
significantly improved the fit (as evidenced by chi square and
residuals), beyond the typical scatter of values seen for the nine different
fits for a given model.
We conservatively adopt 0.04 dex as uncertainty on the derived
luminosities.
Figure~\ref{lenstronomyfits} shows example fits by lenstronomy
with our chosen procedure.
Table~\ref{table:derived} lists all the derived quantities.
Table~\ref{fittingresults} in the appendix gives the details of the
fitting parameters.

Galactic foreground extinction was subtracted based on dust reddening measurements
from \citet{sch11}, assuming F814W = 0.61
$A_V$.
Magnitudes were converted to luminosities, applying a $k$-correction
using PySynphot\footnote{https://pysynphot.readthedocs.io/en/latest/}
\citep{lim15} and a \citet{kin96} Sa galaxy template.
Note that given the low redshift of our galaxies, using a different
template does not significantly change our results.
Also, PySynphot gave (V-I) colors less than 1.2 magnitude
for all galaxy templates (elliptical, S0, Sa, or Sb) and thus,
the F814W filter magnitudes can be considered identical to I-band
magnitudes \citep{har18}.

To derive colors, we fitted the Gemini and SDSS
images (in the filters $g'$, $r'$, $i'$ and $z'$) in a similar way
using lenstronomy.
We first used the Gemini images to independently determine
the host-galaxy morphology of each galaxy, based on both visual
inspection of the images and lenstronomy fitting results.
The conclusions reached are identical to those from the HST fitting.
We then adopted the same host-galaxy parameters derived from the fitting of the HST
images and used lenstronomy to fit the Gemini and SDSS images, leaving only the
magnitudes of PSF and host-galaxy components as free parameters.
This gives us magnitudes in 5 or 6 different filters for the different host-galaxy components
spheroid, (pseudo-) bulge, and disk, if present.
Dust extinction and $k$ correction were applied.

As many literature studies rely on GALFIT, we also 
ran GALFIT on the same background-subtracted HST
images for comparison, using the same PSFs,
error image and general procedure as for lenstronomy.
Overall, the results agree, especially (and not surprisingly) for
all derived values (magnitudes, effective radius and S{\'e}rsic index) for a single-component fit
(spheroid only) and magnitudes, independent of fit used (spheroid,
spheroid-disk or spheroid-disk-bar) \citep[similar conclusions were also reached by][]{yan21}.
  The mean of the ratio between magnitudes as determined
  from lenstronomy and GALFIT is 1.00$\pm$0.01 for spheroid magnitude
  in a spheroid-only fit;
  1.00$\pm$0.03 for spheroid and 1.01$\pm$0.02 for disk 
  in a spheroid-disk fit,
  and 1.02$\pm$0.04 for spheroid, 1.00$\pm$0.06 for disk
  and 1.01$\pm$0.01 for bar in a spheroid-disk-bar fit.
However, for more complicated models,
 the effective radii for individual objects scatter, but the biggest
 difference is seen in the spheroid S{\'e}rsic index (since disk and
 bar have fixed S{\'e}rsic indices) (Fig.~\ref{figure:galfit}). While the mean of the ratio 
 between S{\'e}rsic index $n$ as determined from lenstronomy and GALFIT is still around
 1, it scatters greatly (0.97$\pm$0.6 for spheroid-disk fit and
 0.97$\pm$0.87 for spheroid-disk-bar fit). This cautions the usage of
 $n$ alone as an indication of the nature of the spheroid (classical
 vs. pseudo-bulge) and in this paper, we use a conservative approach
 (see discussion in Section~\ref{host}). However, based on our
 experience using GALFIT in many previous studies \citep[e.g.,][]{ben10,ben11a} we want to stress that GALFIT tends to
 need more user interaction and visual inspections of results to
 ensure a true global minimum was reached which we did not do
 here. Lenstronomy's design of semi-linear inversion and PSO resulted in a
significant improvement in automation and  reduction of
labor-intensive work in the fitting process relative to GALFIT.

\begin{longrotatetable}
\begin{deluxetable*}{lcccccccccc}
\label{table:derived}
\tabletypesize{\footnotesize}
\tablecolumns{11}
\tablecaption{Derived quantities}
\tablehead{\colhead{Object} & 
\colhead{\mbh} &
\colhead{$\sigma_{\rm \star}$}
 & \colhead{$L_{\rm{I, sph}}$}
                 & \colhead{$L_{\rm{I, sph+bar}}$}
                 & \colhead{$L_{\rm{I, host}}$}
                 & \colhead{$M_{\rm sph, dyn}$} 
&                 \colhead{$M_{\rm sph, \star}$} &
  \colhead{$M_{\rm sph+bar, \star}$}
  & \colhead{$M_{\rm host, \star}$}
& \colhead{Host}
                           \\
& (log \msun) &
                                                          (km\,s$^{-1}$)
                                      & (log \msun) & (log \Lsun) &
                                                                    (log
                                                                    \Lsun)
                          & (log \msun) & (log \msun)
  \\
(1) & (2)  & (3) & (4) & (5) & (6) & (7) & (8) & (9) & (10) & (11) 
                                                                     }
                                                                     \startdata
0013-0951 & 8.11 & 109 & 9.48 & ... & 10.38 & 9.55 & 9.89 $\pm$ 0.04 & ... & 10.59 $\pm$ 0.16 & BD (C) \\
0038+0034 & 8.49 & 120 & 10.38 & ... & 10.44 & 10.56 & 10.43 $\pm$ 0.2 & ... & 10.54 $\pm$ 0.18 & BD (C) \\
0109+0059 & 7.78 & 181 & 9.73 & 10.05 & 10.5 & 9.9 & 10.29 $\pm$ 0.01 & 10.38 $\pm$ 0.04 & 10.83 $\pm$ 0.14 & BDB (P) \\
0121-0102 & 8.01 & 133 & 9.23 & 10.02 & 10.74 & 9.73 & ... & ... & ... & BDB (P) \\
0150+0057 & 7.51 & 192 & 9.92 & 10.3 & 10.88 & 10.2 & 10.41 $\pm$ 0.06 & 10.76 $\pm$ 0.11 & 11.15 $\pm$ 0.15 & BDB (P) \\
0206-0017 & 8.26 & 231 & 10.75 & ... & 11.01 & 11.01 & 11.02 $\pm$ 0.16 & ... & 11.25 $\pm$ 0.17 & BD (C) \\
0212+1406 & 7.58 & 175 & 10.03 & 10.11 & 10.66 & 10.28 & 10.49 $\pm$ 0.12 & 10.49 $\pm$ 0.15 & 10.88 $\pm$ 0.18 & BDB (P) \\
0301+0110 & 7.61 & 175 & 10.06 & ... & 10.24 & 10.28 & 10.29 $\pm$ 0.18 & ... $\pm$ ... & 10.33 $\pm$ 0.21 & BD (C) \\
0301+0115 & 7.81 & 97 & 9.66 & 9.97 & 10.38 & 9.33 & 10.13 $\pm$ 0.17 & 10.3 $\pm$ 0.19 & 10.66 $\pm$ 0.2 & BDB (P) \\
0336-0706 & 7.79 & 229 & 10.22 & ... & 10.71 & 10.71 & 10.66 $\pm$ 0.13 & ... & 10.94 $\pm$ 0.18 & BD (P) \\
0353-0623 & 7.76 & 179 & 10.0 & 10.13 & 10.45 & 10.48 & 10.17 $\pm$ 0.18 & 10.33 $\pm$ 0.17 & 10.59 $\pm$ 0.17 & BDB (C) \\
0737+4244 & 7.81 & 179 & 10.0 & ... & 10.54 & 10.48 & 9.93 $\pm$ 0.04 & ... & 10.72 $\pm$ 0.19 & BD (C) \\
0802+3104 & 7.69 & 115 & 9.4 & 9.63 & 10.23 & 9.32 & ... & ... & ... & BDB (P) \\
0811+1739 & 7.43 & 143 & 9.68 & 9.94 & 10.45 & 9.83 & 10.18 $\pm$ 0.11 & 10.35 $\pm$ 0.13 & 10.64 $\pm$ 0.19 & BDB (P) \\
0813+4608 & 7.4 & 124 & 9.96 & 10.18 & 10.38 & 9.87 & 10.32 $\pm$ 0.13 & 10.55 $\pm$ 0.14 & 10.65 $\pm$ 0.15 & BDB (C) \\
0845+3409 & 7.63 & 121 & 9.99 & 10.11 & 10.57 & 10.15 & 10.54 $\pm$ 0.1 & 10.58 $\pm$ 0.12 & 10.8 $\pm$ 0.15 & BDB (C) \\
0857+0528 & 7.68 & 133 & 9.59 & ... & 10.46 & 9.73 & 9.47 $\pm$ 0.18 & ... & 10.55 $\pm$ 0.18 & BD (C) \\
0904+5536 & 8.03 & 133 & 9.68 & ... & 9.95 & 9.73 & 9.93 $\pm$ 0.17 & ... & 9.94 $\pm$ 0.21 & BD (C) \\
0909+1330 & 7.47 & 133 & 9.42 & 9.8 & 10.46 & 9.73 & 9.85 $\pm$ 0.14 & 10.17 $\pm$ 0.15 & 10.55 $\pm$ 0.19 & BDB (P) \\
0921+1017 & 7.71 & 89 & 10.04 & ... & 10.56 & 10.22 & 10.15 $\pm$ 0.18 & ... & 10.64 $\pm$ 0.2 & BD (C) \\
0923+2254 & 7.95 & 89 & 9.94 & 10.1 & 10.63 & 10.22 & 10.04 $\pm$ 0.19 & 10.3 $\pm$ 0.17 & 10.71 $\pm$ 0.2 & BDB (P) \\
0923+2946 & 7.82 & 141 & 10.49 & ... & 10.49 & 10.6 & 10.71 $\pm$ 0.18 & ... & 10.71 $\pm$ 0.18 & B (C) \\
0927+2301 & 7.2 & 193 & 10.02 & ... & 10.66 & 10.23 & 10.41 $\pm$ 0.15 & ... & 10.88 $\pm$ 0.18 & BD (C) \\
0932+0233 & 7.7 & 126 & 9.86 & ... & 10.33 & 9.92 & 10.23 $\pm$ 0.15 & ... & 10.54 $\pm$ 0.17 & BD (C) \\
0936+1014 & 7.85 & 126 & 9.46 & ... & 10.6 & 9.92 & 9.92 $\pm$ 0.08 & ... & 10.83 $\pm$ 0.18 & BD (C) \\
1029+1408 & 8.12 & 185 & 10.53 & ... & 10.66 & 10.83 & 10.79 $\pm$ 0.17 & ... & 10.8 $\pm$ 0.18 & BD (C) \\
1029+2728 & 7.18 & 125 & 9.75 & ... & 10.0 & 9.8 & 9.99 $\pm$ 0.18 & ... & 10.13 $\pm$ 0.19 & BD (C) \\
1029+4019 & 7.94 & 180 & 9.68 & ... & 10.36 & 10.09 & 9.82 $\pm$ 0.07 & ... & 10.52 $\pm$ 0.18 & BD (C) \\
1042+0414 & 7.4 & 64 & 9.51 & 9.72 & 10.15 & 8.87 & 9.81 $\pm$ 0.06 & 10.04 $\pm$ 0.13 & 10.36 $\pm$ 0.17 & BDB (P) \\
1043+1105 & 8.13 & 64 & 9.73 & ... & 9.73 & 8.87 & 9.58 $\pm$ 0.22 & ... & 9.58 $\pm$ 0.22 & B (C) \\
1058+5259 & 7.76 & 122 & 9.99 & 10.23 & 10.46 & 9.92 & 10.21 $\pm$ 0.16 & 10.44 $\pm$ 0.17 & 10.59 $\pm$ 0.18 & BDB (P) \\
1101+1102 & 8.37 & 169 & 10.03 & ... & 10.28 & 10.63 & 10.17 $\pm$ 0.19 & ... & 10.43 $\pm$ 0.17 & BD (C) \\
1104+4334 & 7.3 & 82 & 10.15 & 10.19 & 10.21 & 10.1 & 10.21 $\pm$ 0.19 & 10.25 $\pm$ 0.2 & 10.3 $\pm$ 0.18 & BDB (C) \\
1137+4826 & 7.0 & 150 & 9.59 & ... & 9.88 & 9.6 & 9.81 $\pm$ 0.06 & ... & 10.08 $\pm$ 0.17 & BD (C) \\
1143+5941 & 7.77 & 124 & 9.7 & 10.01 & 10.36 & 9.81 & 10.21 $\pm$ 0.12 & 10.38 $\pm$ 0.14 & 10.46 $\pm$ 0.19 & BDB (C) \\
1144+3653 & 7.99 & 169 & 9.95 & ... & 10.43 & 10.26 & 10.19 $\pm$ 0.17 & ... & 10.56 $\pm$ 0.19 & BD (C) \\
1145+5547 & 7.48 & 169 & 9.28 & 9.61 & 10.49 & 10.26 & 9.73 $\pm$ 0.13 & 10.1 $\pm$ 0.11 & 10.61 $\pm$ 0.19 & BDB (C) \\
1147+0902 & 8.65 & 133 & 10.41 & ... & 10.47 & 10.51 & 10.46 $\pm$ 0.21 & ... & 10.52 $\pm$ 0.2 & BD (C) \\
1205+4959 & 8.26 & 164 & 10.07 & ... & 10.59 & 10.19 & 10.33 $\pm$ 0.17 & ... & 10.71 $\pm$ 0.19 & BD (C) \\
1206+4244 & 7.58 & 164 & 9.72 & 10.05 & 10.56 & 10.19 & 9.97 $\pm$ 0.16 & 10.34 $\pm$ 0.16 & 10.69 $\pm$ 0.19 & BDB (P) \\
1216+5049 & 8.32 & 165 & 10.07 & ... & 10.39 & 10.51 & 10.3 $\pm$ 0.17 & ... & 10.59 $\pm$ 0.18 & BD (C) \\
1223+0240 & 7.36 & 165 & 9.71 & ... & 10.08 & 10.51 & 9.9 $\pm$ 0.18 & ... & 10.16 $\pm$ 0.2 & BD (C) \\
1246+5134 & 7.19 & 105 & 9.6 & ... & 10.07 & 9.59 & 10.08 $\pm$ 0.08 & ... & 10.33 $\pm$ 0.17 & BD (P) \\
1306+4552 & 7.42 & 89 & 9.2 & 9.79 & 10.39 & 9.22 & ... & ... & ... & BDB (P) \\
1307+0952 & 7.48 & 89 & 9.53 & 9.59 & 10.44 & 9.22 & 9.88 $\pm$ 0.02 & 9.88 $\pm$ 0.15 & 10.56 $\pm$ 0.18 & BDB (P) \\
1312+2628 & 7.77 & 89 & 9.5 & 9.8 & 10.54 & 9.22 & 9.96 $\pm$ 0.02 & 10.24 $\pm$ 0.13 & 10.57 $\pm$ 0.2 & BDB (P) \\
1405-0259 & 7.3 & 124 & 9.65 & ... & 10.44 & 9.94 & 10.06 $\pm$ 0.13 & ... & 10.64 $\pm$ 0.18 & BD (C) \\
1416+0137 & 7.52 & 149 & 10.2 & ... & 10.66 & 10.41 & 10.61 $\pm$ 0.14 & ... & 10.73 $\pm$ 0.19 & BD (C) \\
1419+0754 & 8.26 & 214 & 10.16 & ... & 10.89 & 10.53 & 10.61 $\pm$ 0.13 & ... & 11.06 $\pm$ 0.18 & BD (P) \\
1434+4839 & 7.92 & 119 & 9.71 & 9.99 & 10.43 & 9.8 & 9.76 $\pm$ 0.18 & 10.13 $\pm$ 0.18 & 10.48 $\pm$ 0.2 & BDB (P) \\
1545+1709 & 8.29 & 165 & 10.01 & ... & 10.19 & 10.42 & 10.04 $\pm$ 0.2 & ... & 10.29 $\pm$ 0.18 & BD (C) \\
1557+0830 & 7.92 & 165 & 9.74 & ... & 9.74 & 10.42 & 9.71 $\pm$ 0.21 & ... & 9.71 $\pm$ 0.21 & B (C) \\
1605+3305 & 8.08 & 200 & 9.8 & ... & 10.2 & 10.29 & 9.9 $\pm$ 0.19 & ... & 10.4 $\pm$ 0.17 & BD (C) \\
1606+3324 & 7.8 & 162 & 10.13 & ... & 10.38 & 10.38 & 10.46 $\pm$ 0.16 & ... & 10.59 $\pm$ 0.18 & BD (C) \\
1611+5211 & 7.93 & 116 & 9.98 & ... & 10.22 & 9.76 & 10.23 $\pm$ 0.16 & ... & 10.38 $\pm$ 0.18 & BD (C) \\
1636+4202 & 8.12 & 149 & 10.34 & 10.38 & 10.49 & 10.74 & 10.44 $\pm$ 0.18 & 10.5 $\pm$ 0.18 & 10.62 $\pm$ 0.19 & BDB (C) \\
1708+2153 & 8.46 & 167 & 10.23 & ... & 10.57 & 10.62 & 10.62 $\pm$ 0.13 & ... & 10.76 $\pm$ 0.2 & BD (P) \\
2116+1102 & 8.25 & 167 & 9.92 & 10.02 & 10.67 & 10.62 & 10.14 $\pm$ 0.16 & 10.32 $\pm$ 0.16 & 10.74 $\pm$ 0.19 & BDB (P) \\
2140+0025 & 7.78 & 113 & 10.05 & ... & 10.48 & 9.82 & 10.33 $\pm$ 0.11 & ... & 10.69 $\pm$ 0.15 & BD (C) \\
2215-0036 & 7.87 & 113 & 9.91 & ... & 10.58 & 9.82 & 10.04 $\pm$ 0.17 & ... & 10.83 $\pm$ 0.18 & BD (C) \\
2221-0906 & 8.03 & 97 & 10.0 & ... & 10.38 & 9.98 & 10.05 $\pm$ 0.15 & ... & 10.66 $\pm$ 0.16 & BD (C) \\
2222-0819 & 7.92 & 107 & 9.48 & 10.0 & 10.79 & 9.61 & 8.29 $\pm$ 0.04 & 10.08 $\pm$ 0.22 & 10.96 $\pm$ 0.22 & BDB (P) \\
2233+1312 & 8.37 & 207 & 9.93 & 10.32 & 10.87 & 10.3 & 10.54 $\pm$ 0.06 & 10.84 $\pm$ 0.11 & 11.16 $\pm$ 0.15 & BDB (P) \\
2254+0046 & 7.63 & 207 & 10.1 & ... & 10.36 & 10.3 & 10.24 $\pm$ 0.12 & ... & 10.51 $\pm$ 0.18 & BD (C) \\
2327+1524 & 7.78 & 261 & 10.56 & ... & 10.85 & 10.86 & 11.01 $\pm$ 0.12 & ... & 11.18 $\pm$ 0.15 & BD (C) \\
2351+1552 & 8.34 & 179 & 10.28 & ... & 10.61 & 10.49 & 10.59 $\pm$ 0.13 & ... & 10.91 $\pm$ 0.14 & BD (C) \\
\enddata
  \tablecomments{
Col. (1): Target ID used throughout the text (based on R.A. and declination). 		      	   	  
Col. (2): Logarithm of \mbh~(solar units) (uncertainty of 0.4 dex).
Col. (3): Stellar-velocity dispersion within
spheroid effective radius (in km\,s$^{-1}$) (uncertainty of 0.04 dex).
Col. (4): Logarithm of spheroid I-band luminosity (solar units)
(uncertainty of 0.04 dex).
Col. (5): Logarithm of spheroid+bar I-band luminosity (solar units)
(uncertainty of 0.04 dex).
Col. (6): Logarithm of host I-band luminosity (solar units)
(uncertainty of 0.04 dex).
Col. (7): Logarithm of spheroid dynamical mass (solar units) (uncertainty of 0.1
dex).
Col. (8):  Logarithm of spheroid stellar mass (solar units).
Col. (9): Logarithm of spheroid+bar stellar mass (solar units).
Col. (10): Logarithm of host stellar mass (solar units).
Col. (11): Host-galaxy fit (B: spheroid only, BD: spheroid+disk, BDB:
spheroid+disk+bar). In parentheses: Spheroid component: C = classical
bulge; P = pseudo-bulge.
}
\end{deluxetable*}
\end{longrotatetable}

\subsection{Stellar-velocity dispersion and black hole mass}
\label{mbh}
In the literature, the observed correlation between \mbh~and $\sigma$ is
generally considered the tightest and thus most fundamental of the \mbh-host-galaxy
scaling relations \citep{tre02,bei12,sag16,sha16,van16,den19}.
Moreover, it is used to calibrate \mbh~by matching the
\mbh-\s~relation of RM AGNs to that of quiescent galaxies.
Thus, robust measurements of \s~are essential. There are several
definitions of $\sigma$ used in the literature, resulting in widely
differing measurements depending on aperture size used and host-galaxy
morphology \citep[see paper III of this series][]{ben15}, with the
most robust being spatially-resolved stellar-velocity dispersions
within the effective spheroid radius. 
Spatially-resolved stellar-velocity dispersions were presented 
in paper II \citep{har12} based on our Keck spectra.
In paper III \citep{ben15}, we determined $\sigma$
from spatially-resolved $\sigma$
measurements integrated within the effective spheroid radius
\citep[see equation (1) in paper III;][]{ben15}.
However, the effective spheroid radius in paper III was based on
surface-photometry of SDSS images.
We here repeat the same calculation, now using robust effective spheroid radii from
the HST surface photometry.
When compared, on average, the $\sigma$ values are similar
(the ones based on HST radii are larger by 1\%),
but with a large scatter of 10\%,
and a couple of individual objects having changed by as much as 50\%.
For 16 objects, the lack of sufficient spatially-resolved $\sigma$
measurements hindered a robust determination of $\sigma$ within the
effective spheroid radius and they were excluded here. Thus, the
\mbh-\s~relation presented in section~\ref{relations} includes 50 objects.

\mbh~was determined for the entire sample in
\citet{ben15},
based on the second moment of the broad H$\beta$ emission line
determined from Keck spectroscopy. The 5100\AA~AGN luminosity was
used as a proxy for BLR size and combined with the width of H$\beta$
to estimate \mbh~as
in equation (2) in \citet{ben15}. In \citet{ben15},
a virial factor of $\log f = 0.71$ was assumed
\citep{par12,woo15}. However, since this virial factor is based on
matching the RM AGN sample to a sample of quiescent galaxies from \citet{mcc11},
we here derived $f$ independently by matching the 
\mbh-\s~relation to that of \citet{kor13}. To do so,
we first fix the slope of the \mbh-\s~relation to the one from
\citet{kor13} and then adjust $\log f$ to match the intercept,
resulting in $\log f = 0.97$.
A wide spread in virial factors ($\log f$ ranging between 0.5 and 1.2)
has also been found in previous studies, 
depending on the choice of different quiescent samples, fitting
methods and \mbh~range \citep[e.g.,][]{par12,sha19}, possibly
due to selection effects in the local sample of quiescent black holes.

\subsection{Stellar and dynamical masses}
From our surface photometry (Section~\ref{surface}), we have magnitudes 
for five to six different bands (HST UVIS/F814W, SDSS $g'$, $r'$,
$i'$, $z'$ for all objects plus Gemini NIRI/Ks for 14 objects)
for the different host-galaxy components, (pseudo-) bulge, disk,
and bar, if present.
To estimate stellar masses from colors, we use a Bayesian stellar-mass estimation
code with priors on age, metallicity and dust content of
the galaxy and error bars on the different magnitudes
\citep{aug09}. To explore the full parameter space and quantify
degeneracies, it uses an MCMC sampler. A Chabrier initial mass function (IMF)
 was assumed, but later, for comparison with literature, converted to
 a Kroupa IMF (by adding 0.075 to $\log M$).
 This gives us stellar masses for the different host-galaxy
 components of 63 objects; for 3 objects, no robust stellar masses
 could be determined.
 Thus, the \mbh-stellar-mass relations presented in section~\ref{relations}
 include 63 objects.
  (Note that the difference in the stellar-mass estimates based on
   HST+SDSS vs. HST+SDSS+Gemini is within the uncertainties of the
   stellar-mass estimates in the HST+SDSS measurements ($\sim$
   0.15-0.2 dex).
In particular, no bias is introduced when adding the Gemini results.
However, the uncertainties in the stellar-mass measurements are
smaller when the K-band magnitudes are included (Table~\ref{table:derived}), so we
use the K-band magnitudes for the stellar-mass determination when available.)

Given $\sigma$ within the effective radius as described in the
previous section, we can also calculate a dynamical mass:
\begin{eqnarray}
  M_{\rm sph, dyn} = c r_{\rm eff, sph} \sigma^2_{\rm ap, reff} / G
  \end{eqnarray}
  with $c$ = 3 for comparison with literature
  \citep{cou14}.
  Since robust $\sigma$ measurements within the effective spheroid
  radius were only obtained for 50 objects, the \mbh-$M_{\rm sph,
    dyn}$ relation in section~\ref{relations} includes 50 objects.

\subsection{Comparison samples}
\label{sec:comparisonsample}
To compare the resulting scaling relations of \mbh~and \s, 
luminosity and mass with the literature, we use
the sample presented by 
\citet{kor13} as a
quiescent galaxy comparison sample, 85 local galaxies with \mbh~based on dynamical modeling of spatially-resolved kinematics.
Their sample consist of 44 elliptical galaxies, 20 spiral and S0 galaxies with
classical bulge and 21 spiral and S0 galaxies with pseudo-bulge.
Five of the elliptical galaxies are mergers in progress.
Pseudo-bulges and mergers are significant outliers in \citet{kor13}
and ignored here.
For 11 objects, the \mbh~is considered uncertain and these objects are
also ignored. We are thus left with 51 objects total, 32 elliptical
galaxies and 19 spiral and S0 galaxies with classical bulges.

The stellar-velocity dispersions are adopted in most cases from
\citet{gue09} and represent effective velocity dispersions within
$r_{\rm eff}$/2 as average of $V^2(r)+\sigma^2(r)$ weighted by $I(r)
dr$, thus consistent with the way we derived stellar-velocity
dispersions, since the difference between
averaging inside $r_{\rm eff}$ and $r_{\rm eff}$/2 is small \citep{kor13}.

\citet{kor13} list spheroid magnitudes in Ks and V, and
(V-Ks) and (B-V) colors.
We use a variety of elliptical and spiral spectral templates from \citet{bru03}
and \citet{kin96} 
and derive a linear least-square fit of the form
$(V-I)=\alpha * (B-V)+ \beta$ with $\alpha=0.72$
and $\beta=0.41$, for conversion to I-band magnitudes.

The stellar spheroid masses given by \citet{kor13} are derived from a mean of mass-to-light
ratios based on $\sigma$ and $(B-V)_0$ (their equations 8 and 9) and
K-band magnitude. The mass-to-light ratio based on color is derived
from \citet{int13}, who assume a Kroupa (2001) IMF, but \citet{kor13}
adjust to the dynamical zeropoint.

For the \mbh-\s~relation, we also show the 29 RM AGN sample presented by \citet{woo15}.
We adjust their \mbh~to match the virial factor of $\log f = 0.97$
adopted here.

\section{RESULTS AND DISCUSSION}
\label{results}
\subsection{Host-galaxy morphology}
\label{host}
The host-galaxy morphology was determined based on visual inspection
of images and the results of the surface-brightness fitting (Section~\ref{surface}).
Of the full sample of 66 AGNs with HST images, we conclude that 3 
are hosted by bona-fide elliptical galaxies and 63 by spiral or S0
galaxies. Out of the latter, 26 galaxies are found to have a bar.
Four objects show signs of interaction and/or merger activity
(0206$-$0017, 0904+5536, 1708+2153, 2254+0046).
The distribution of host-galaxy morphologies is typical for
Seyfert-type AGNs.

In order for a spheroidal component to be classified as pseudo-bulge,
we conservatively require that at least three of the following four
criteria are met \citep[following][]{kor13}:
(i) S{\'e}rsic index $<$ 2;
(ii) bulge-to-total luminosity ratio $<$ 0.5;
(iii) rotation dominated, i.e., ratio between maximum rotational
velocity at effective spheroid radius and central stellar-velocity
dispersion $>$ 1;
(iv) for face-on galaxies, the presence of a bar is considered an
indicator for the existence of a pseudo-bulge.

Table~\ref{table:sample} gives the host galaxy classification for all
objects, including whether or not the above four criteria are
met. \citet{kor13} argue that if the bulge-to-total luminosity
ratio is $>$ 0.5, the bulge can be considered a classical bulge.
As can be seen from the table, all objects that have a bulge-to-total luminosity
ratio $>$ 0.5 were indeed classified as a classical bulge,
based on our conservative requirement above.

In this way, of the 63 spiral or S0 galaxies,
22 spheroids are classified as pseudo-bulges,
the majority of which (19) are in barred spiral galaxies.
Given the wide range of host-galaxy morphologies (with mass-to-light ratios ranging between
0.7 and 2.2) and the high quality of the imaging,
our sample is ideal to study dependency of the \mbh-host-galaxy
scaling relations with other parameters such as (pseudo) bulges and
bars.

Note that this classification is conservative and may
  underestimate the true fraction of pseudo-bulges:
 for face-on galaxies, the true rotation cannot be
  reliably measured, for edge-on galaxies, bars can easily be missed.
  To estimate the fraction of potentially mis-classified bulges
  (classical instead of pseudo-bulge), we carefully inspected all images.
  We consider ten galaxies (0013-0951, 0038+0034, 0121-0102,
  0150+0057, 0301+0110, 0921+1017, 1205+4959, 1312+2628, 2116+1102,
  2222-0819) as face-on (or close to face-on) galaxies.
  Out of these, five galaxies are already classified as
  pseudo-bulges based on meeting three of the four criteria (0121-0102,
  0150+0057, 1312+2628, 2116+1102, and 2222-0819).
  For four galaxies, either no criterion is met (0038+0034) or only one criterion (0301+0110, 0921+1017,
  1205+4959), so their classifications as classical bulge is
  independent of criterion (iii). There is only one object 
  (0013-0951) that meets two criteria, but does not seem
  rotation-dominated and this could indeed be a pseudo-bulge.
  As for highly-inclined galaxies, there are six we consider in more detail
  (0336-0706, 0936+1014, 1216+5049, 1246+5134,  1545+1709, 2215-0036).
  One already meets three criteria (1246+5134) and is already
  classified as a pseudo-bulge. Two other objects either meet no
  criterion (1545+1709) or  only meet one
  criterion (1216+5049). That leaves three objects that may
  potentially be mis-classified as classical bulges 
  (0336-0706, 0936+1014, 2215-0036).
  Thus, the number of pseudo-bulges in the sample can be considered a lower limit.
  However, as we discuss in the next section, there are no
  outliers (whether pseudo-bulge or classical bulge) in the scaling relations, so the conclusions remain unchanged.

 \subsection{Scaling relations}
\label{relations}
\begin{figure*}
  \center
    \includegraphics[scale=0.35]{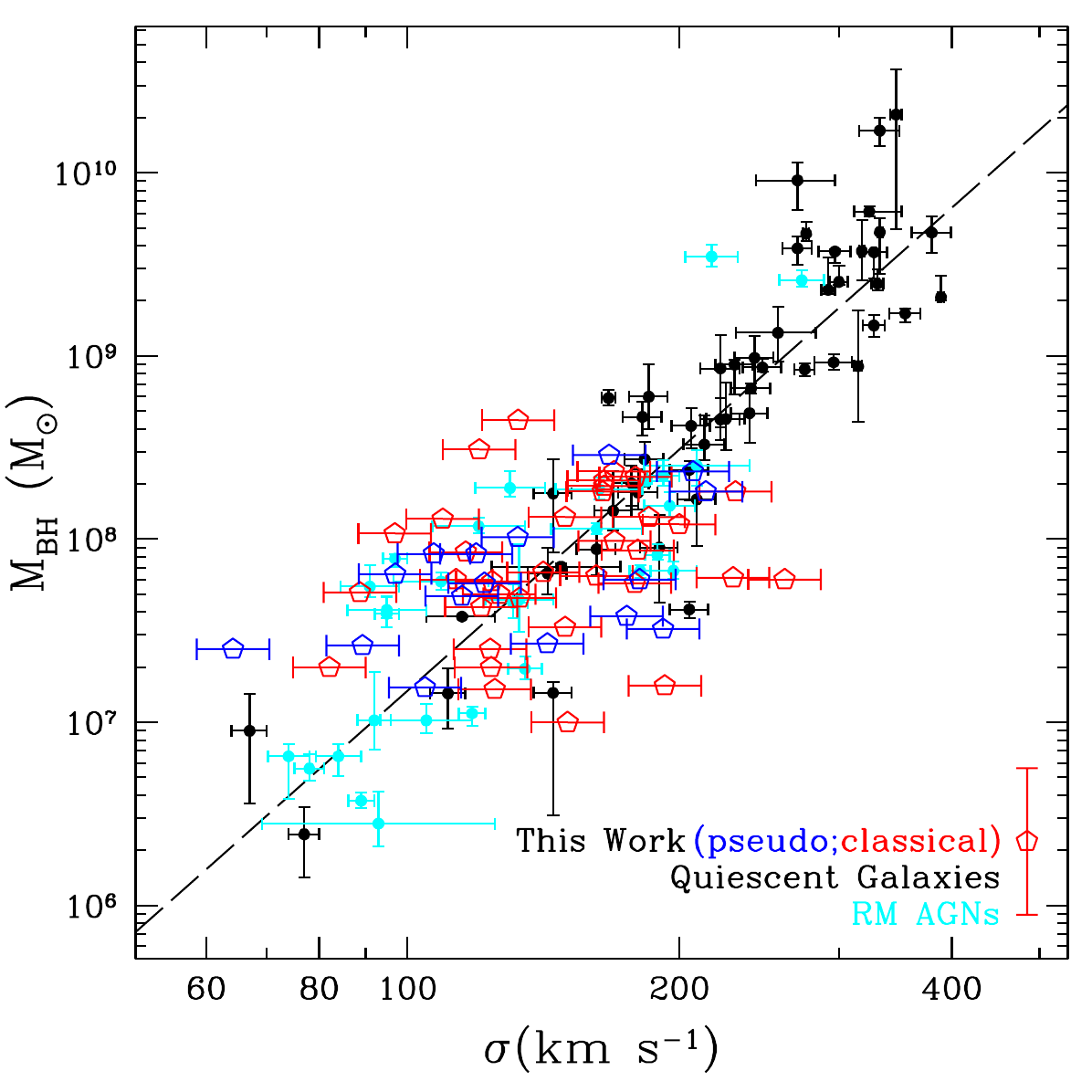}
    \includegraphics[scale=0.35]{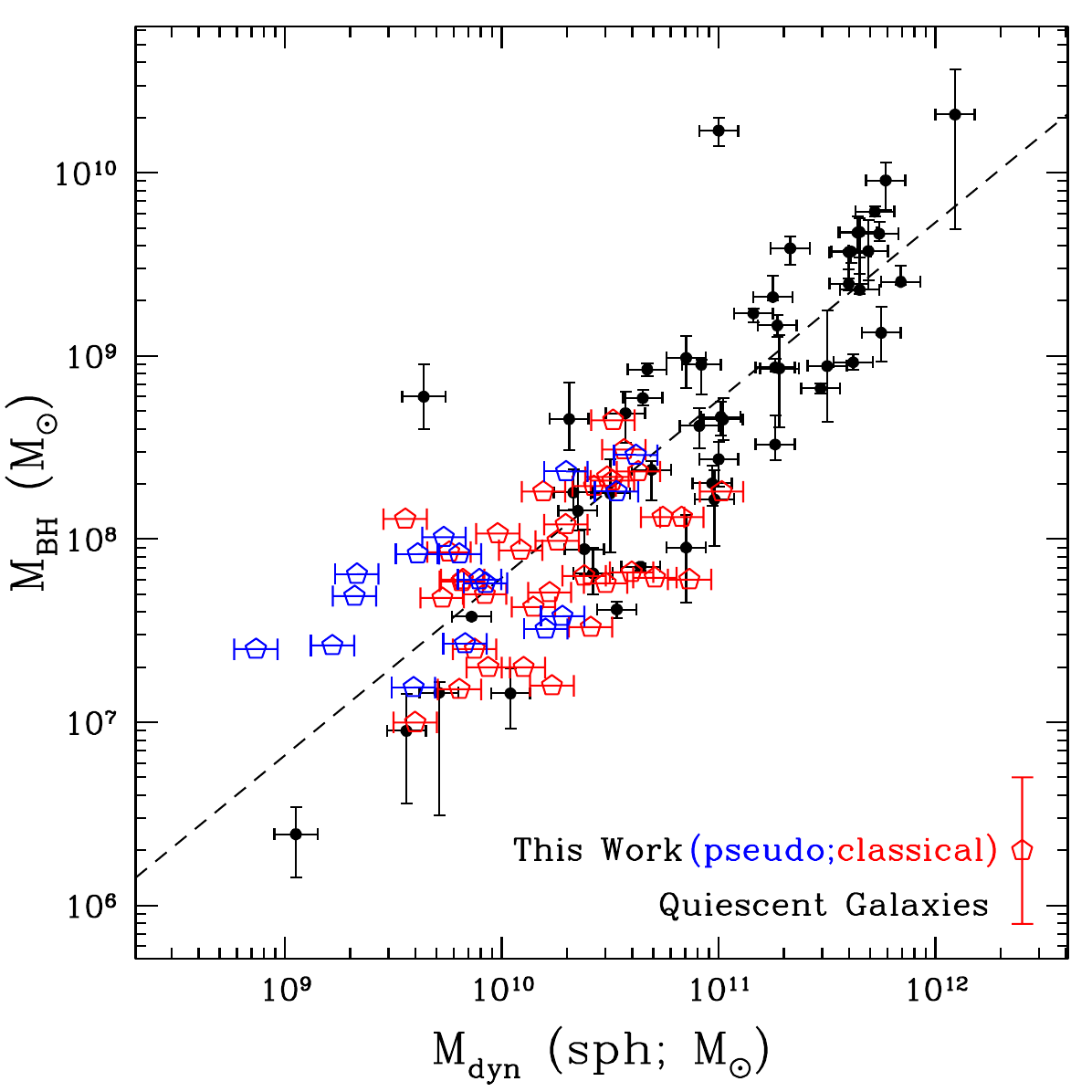}
    \includegraphics[scale=0.35]{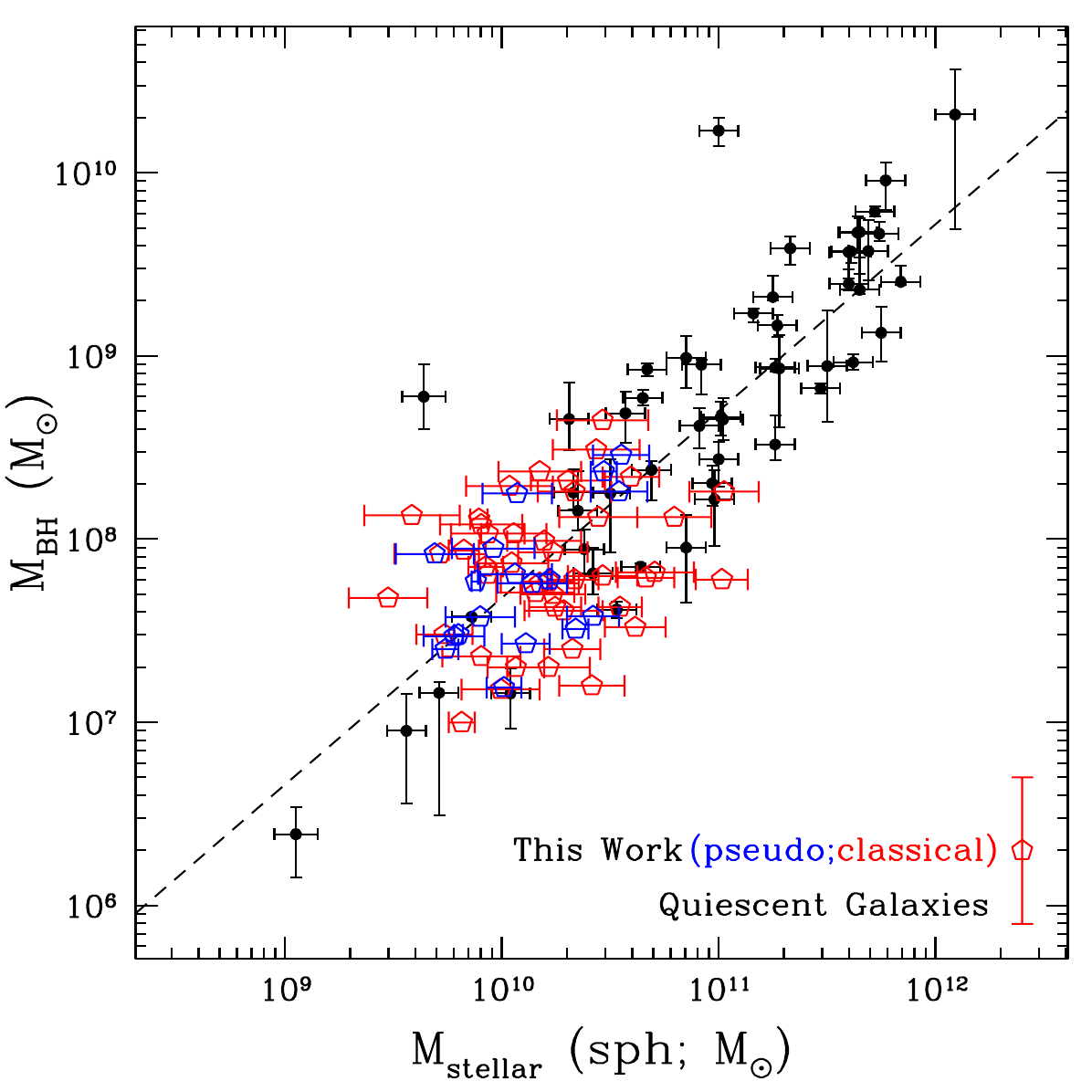}
  \includegraphics[scale=0.35]{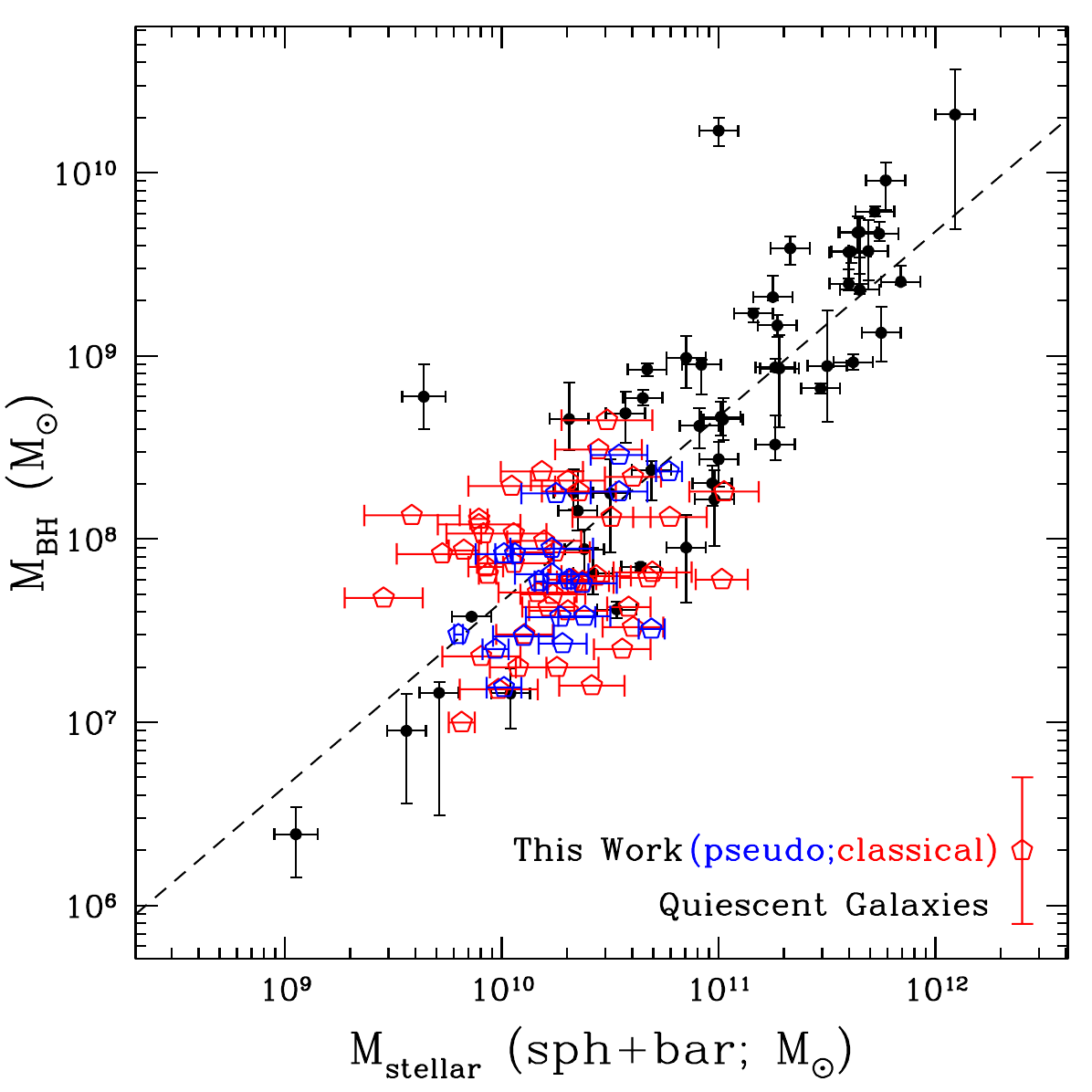}
  \includegraphics[scale=0.35]{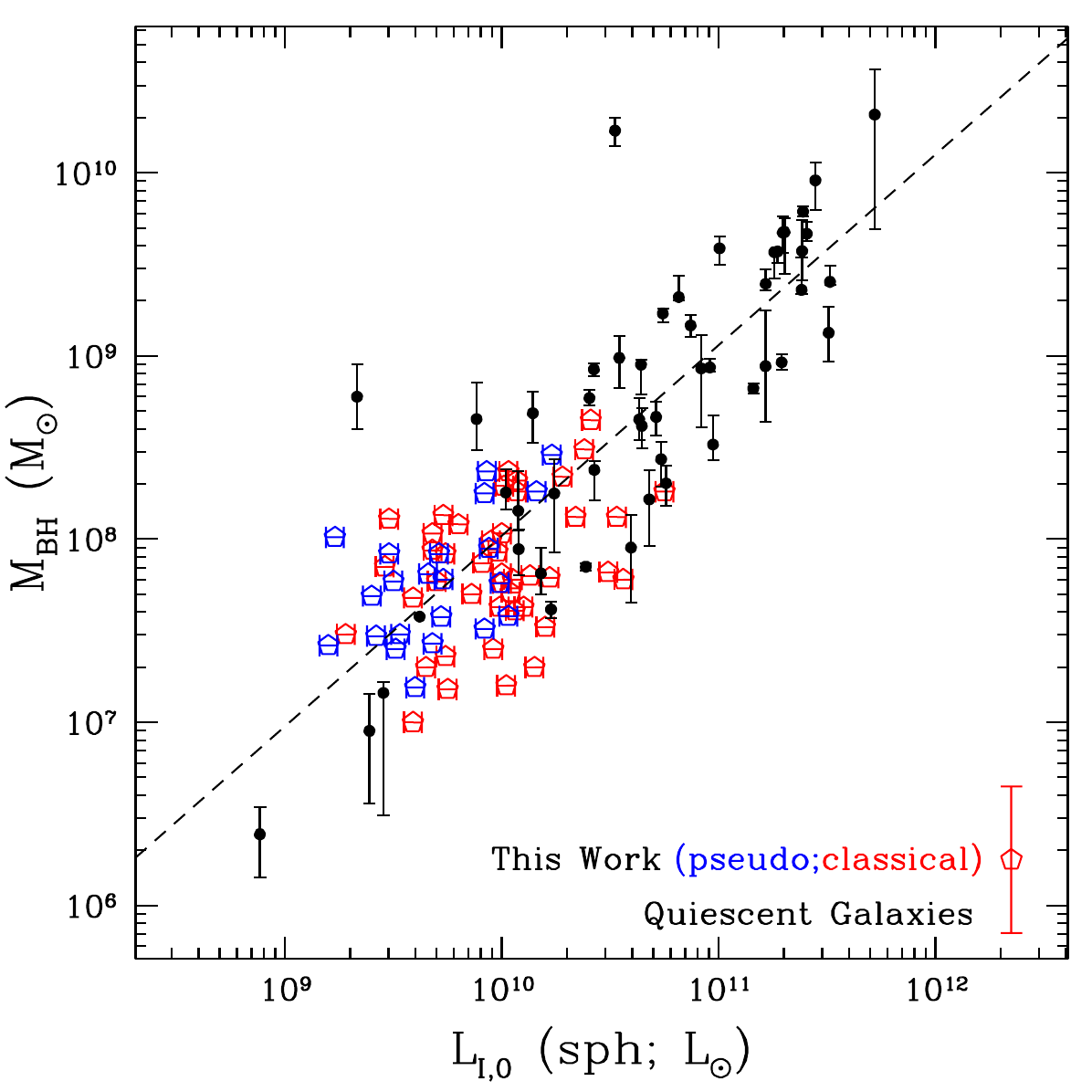}
  \includegraphics[scale=0.35]{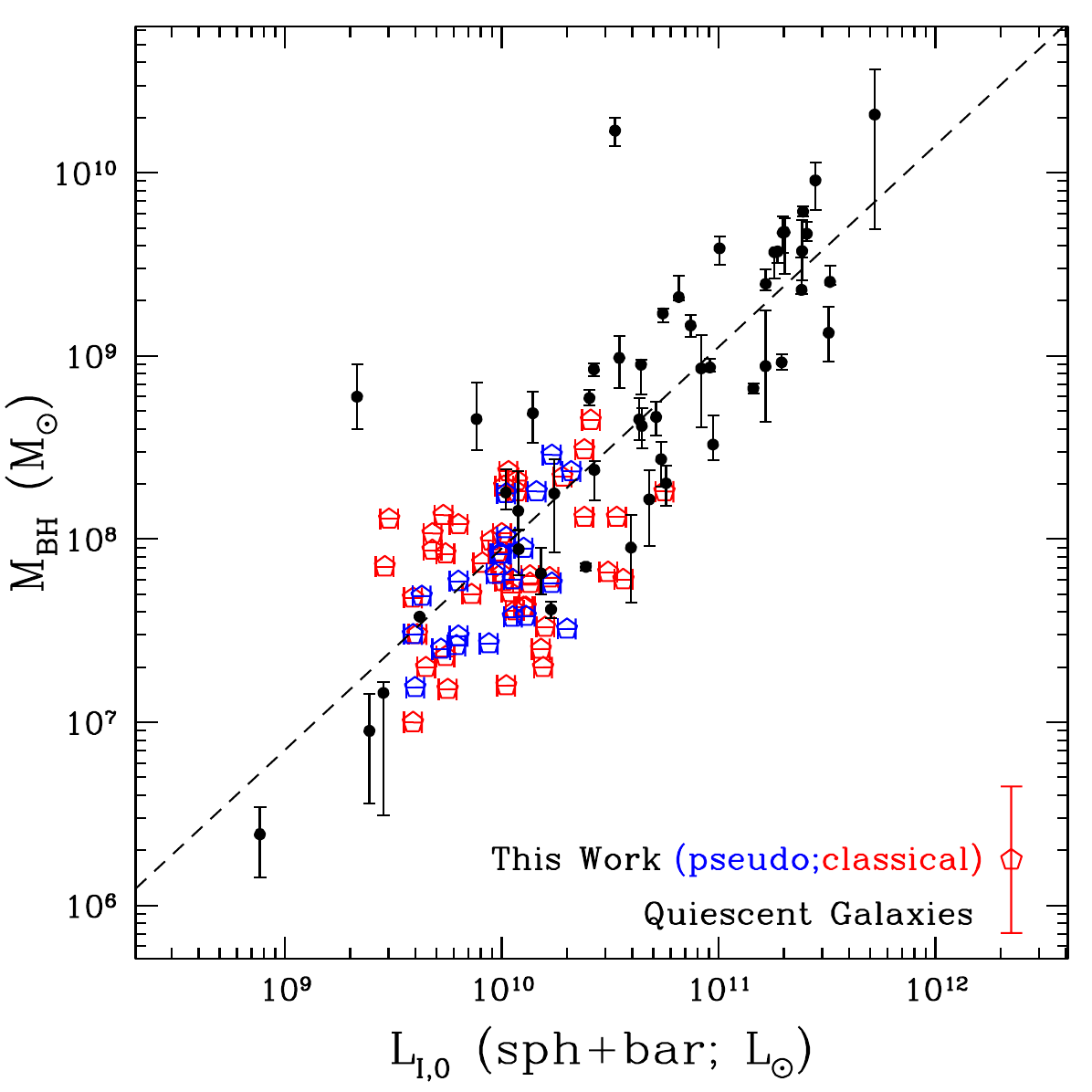}\\
  \caption{\mbh~scaling relations. In 
    all panels, black data points correspond to the local quiescent
    comparison sample from \citet{kor13}, only including elliptical
    galaxies and spiral galaxies with classical bulge.
    For our sample, pseudo-bulges are shown in blue and classical
    bulges in red. To reduce confusion of data points, error bars on \mbh~for our sample
are omitted and shown instead in the bottom right corner.
    Top left panel: \mbh-\s~relation. Cyan data points show 29 RM
    AGNs from \citet{woo15}.
    Top right panel: \mbh-$M_{\rm sph, dyn}$ relation.
    Middle left panel: \mbh-$M_{\rm sph}$ relation.
    Middle right panel: \mbh-$M_{\rm sph+bar}$ relation.
    Bottom left panel: \mbh-$L_{\rm sph, I}$ relation.
    Bottom right panel: \mbh-$L_{\rm sph+bar, I}$ relation.
    \label{figure:mbh}}
\end{figure*}

We present scaling relations between \mbh~and stellar
velocity dispersion $\sigma$ (within effective radius of spheroid),
dynamical spheroid mass, stellar mass, and I-band luminosity (Figure~\ref{figure:mbh}).
We choose \citet{kor13} for a consistent comparison for all these
different scaling relations, even though there
are more recent studies with a compilation of larger
samples. However, in a review by \citet{gre20},
the authors note that their results on the \mbh-$\sigma$ relation
would not have changed 
if, instead of using a recent literature compilation, they had used exclusively the \citet{kor13} sample.
For the \mbh-$\sigma$ relation, we also show 29 RM AGNs
measured in a similar way \citep[][see Section \ref{sec:comparisonsample} for details]{woo15}.

Following common practices, we fit the different scaling relations as
the linear relation with $\alpha$ and $\beta$ as slope and
intercept values (Table~\ref{table:fits}). The error bars of the
measurements (in both x and y) are taken into account to perform the
inference. We first use SciPy \citep{vir20} to estimate the
  minimization of these parameters. Then, we use the minimization results
  as initial  values to run MCMC \citep[using the EMCEE package][]{for13}.
  After burn-in, the MCMC chain median values are adopted as the
  ‘best-fit’ values which are presented in the paper.
  The upper and lower limits are the 84\% and 16\% of the chain histogram, respectively.
The intrinsic scatter is estimated so that when the squares of the
observed uncertainties are summed up, the best-fit reduced chi-square value is close to unity.
We fit all samples (including RM AGNs and quiescent galaxies) using
  the same code, for consistency, rather than using literature values. However, none of our conclusions would change
  if we were to use the fits given in the literature instead \citep{kor13,woo15}.

For all of the \mbh~scaling relations, our sample of 66 local AGNs
naturally extends the correlations for quiescent galaxies down to
\mbh$\sim10^7$ M$_\odot$ along
the same line, with the same slope and normalization.
However, by itself, the dynamic range in \mbh~covered by our sample
is too small to determine the slope. Thus, when deriving fits to the different scaling
relations, we either fit both samples (AGNs and quiescent galaxies)
together or when fitting our AGN sample alone, we fix the slope to that of \citet{kor13}.

The \mbh-$\sigma$ relation of our local AGN sample
with \mbh~determined using the single-epoch method
and $\sigma$ based on spatially-resolved kinematics
agrees with that of AGNs with \mbh~obtained from reverberation
mapping \citep{par12,woo15}.
The importance of the RM
AGN sample cannot be overstated since it serves as the \mbh\ calibrator beyond the local Universe.
Given that, within the uncertainties, slope and scatter of the \mbh-$\sigma$ relation of our local AGNs, selected based solely on broad H$\beta$ emission
line width, agree with that of RM AGNs provides a
confirmation that the selection of the RM AGN sample based on
variability (not on well defined galaxy/black-hole mass properties)
does not introduce biases.
Moreover, the close agreement between both samples provides
an indirect validation of the single-epoch method for the estimation
of \mbh.
(Note that these conclusions are independent of the fact that
  the \mbh-\s~scaling relation of our local AGNs  is matched to that of inactive galaxies,
  since that only affects the normalization, but not slope and scatter.
  Moreover, we do not determine a virial factor for the RM
  AGNs separately, but use the same one as for our local AGNs,
  resulting in a good agreement, providing an additional check
  that the two samples match.)

To illustrate the effect of un-identified bars,
we also include scaling relations for stellar mass and luminosity
with spheroid+bar component added.
This may help in comparison with literature data,
especially given the difficulties and potential ambiguities involved
in decomposition of images with poor data quality.
Since the spheroids in the majority of barred spiral galaxies in our
sample are classified as pseudo-bulges (19/26),
this affects the location of pseudo-bulges the most.
It moves the pseudo-bulges further to the right in the \mbh-stellar
mass and \mbh-luminosity relations which tends to move them into
better agreement with the scaling relations of quiescent galaxies.
However, within the uncertainty, the difference is small.
For none of the scaling relations do we find a significant
difference  between pseudo- and classical bulges in terms of
 correlations with \mbh.
This is in line with some studies \citep[e.g.,][]{dav18},
but contrary to many others \citep[e.g.,][]{hu08,
  gre10,sani11,lae16,sag16,men16,den19}. For example,
\citet{kor13} went so far to conclude that ``any \mbh~correlations with the properties of
disk-grown pseudobulges [...] are weak enough to imply no close
coevolution'' \citep[see also,][]{kor11}.

Pseudo-bulges, considered to have evolved secularly through
dissipative processes rather than through
galaxy mergers \citep[e.g.,][]{cou96,kor04},
play an important role for understanding the origin of the
\mbh~scaling relations.
If major mergers are the fundamental drivers of the \mbh~scaling
relations, only classical bulges, centrally concentrated, mostly red and quiescent,
merger-induced systems, should follow these tight correlations.
On the basis of high-quality HST imaging, a careful 
analysis and a conservative classification of bulges as pseudo-bulges,
our results clearly show that pseudo-bulges follow the same relations as classical
bulges, confirming findings of an earlier study of ours based on
SDSS images \citep{ben11a}.
This is in line with studies that argue
  that most of the growth of the SMBH
  happens gradually over time via secular processes
  \citep[e.g.,][]{sim17,mar18}.
(Note that none of our results would change if we
excluded the pseudo-bulges from our sample, to more closely match the
comparison sample by \citet{kor13} for which we excluded pseudo-bulges.
While our sample consists of a significant fraction of pseudo-bulges
(22 of 66 galaxies), when repeating all the fits presented  in
Table~\ref{table:fits} without pseudo-bulges, we obtain fits with the
same parameters within the uncertainties. This is not surprising and
highlights the fact that pseudo-bulges do not form outliers in our sample.)

In fact, our study shows that there are no
significant outliers that could be attributed to any specific
category, whether it be galaxies with pseudo-bulges, bars or signs of interactions/mergers.
For example, the four objects with signs of mergers/interaction do not
tend to lie off the relations.
Likewise, barred galaxies (26 out of 63 disk galaxies in our sample) do not form
outliers, in line with some literature
\citep{bei12,sahu19}.
The location of barred galaxies on the scaling relations is not only
important since over half of the disk galaxy population is barred
\citep[e.g.,][]{wei09}, but also because of the relevance of bars in
secular evolution and potentially fueling of BHs.
Moreover, it is much easier to identify a bar than a pseudo-bulge
\citep[for a discussion, see][]{gra16}, a reason why some studies
choose to distinguish between barred and non-barred galaxies
rather than classical vs. pseudo-bulges \citep[e.g.,][]{gra13}.
Most previous literature studies found barred galaxies to lie off
the \mbh~scaling relations, in particular in the case of \mbh-\s~
\citep[e.g.,][for conflicting results, see also
\citealt{bei12}]{gra08,gra09,hu08}.
This is not surprising, given that the stellar dynamics in
galactic sub-structures such as bars and pseudo-bulges
is very different from that of elliptical galaxies or classical
bulges. Moreover, $\sigma$ measurements can depend significantly on, e.g., size of the
fiber (as is the case, for example, for SDSS), orientation of the slit (in case of
long-slit observations) and aperture size used 
\citep[for details and comparisons, see ][]{ben15}.
Integral-field spectroscopy and spatially-resolved spectroscopy is an
obvious step forward and has been obtained for a sub-sample of the RM
AGN sample \citep{bat17}.
While our $\sigma$ measurements were obtained using long-slit
spectroscopy, we mitigate these effects by using spatially-resolved measurements integrated
within the spheroid effective radius which is a robust way to
determine $\sigma$ \citep[see also][]{ben15}.
(Note that while we may be underestimating the true fraction of
  barred galaxies in our sample, given they are hard to identify in
  edge-on galaxies (see Section~\ref{host}), this does not affect our conclusions,
  since there are no outliers (barred or not) in the scaling relations.)

The majority of AGNs in our sample reside in host
galaxies of S0 or late-type morphology (63/66), out of which
almost half of the galaxies are barred and a third of
spheroids are classified as pseudo-bulges.
The fact that all of them are obeying
the same tight \mbh~scaling relations,
highlights the importance of
secular evolution for the growth of BHs and bulges.
Secular evolution may have a synchronizing effect,
growing BHs and bulges simultaneously
at a small but steady rate for late-type galaxies,
and keeping them on tight relations over time.
Comparison with semi-analytical models for galaxy formation
including secular evolution \citep[such as e.g.,][who, however,
find little or no correlation of pseudo-bulge mass with \mbh]{men16} can further shed light on such a scenario,
but is beyond the scope of this paper.

The \mbh-$\sigma$ relation is considered 
the most fundamental of the scaling relations,
due to its tightness (0.3 dex in $\log$ \mbh)
\citep{tre02,bei12,sag16,van16,den19},
or based on residuals and principle-component analyses
\citep{sha16,mar21},
at least for late-type
spiral galaxies \citep[][see, however,
\citealt{dav19}]{gue09,gre10,lae16}.
Interestingly, we do not find significant differences in the tightness
of the different correlations. The scatter in the relations ranges between
0.2-0.4 dex, smaller or equal to that of quiescent galaxies
\citep{gue09, mcc13,kor13}.
(Note that even though 
the quiescent galaxies span a larger range in
\mbh, we can still compare the scatter between the different samples,
assuming that the scatter is independent of \mbh.)
We attribute this difference to a combination of  (i) 
our homogeneous sample selection, (ii) high-quality data, for both imaging and
spectroscopy, and (iii) reliable surface photometry
for a detailed structural decomposition of the host galaxy components and
spatially-resolved kinematics.
Given the fact that the biggest uncertainty in host-galaxy
surface-brightness fitting is the classification and identification
of individual structures, a combination of (ii) and (iii) is
essential if one wants to determine the role of host-galaxy
sub-structures on the correlation with \mbh.

\section{SUMMARY}
\label{summary}
This paper presents a study of 66 local ($0.02 \le z \le 0.1$) active galactic nuclei (AGNs)
homogeneously selected based on the presence of a broad H$\beta$
emission line in SDSS spectra. High-quality HST optical (66 objects) and
Gemini NIR imaging (14 of 66 objects) are complemented by spatially-resolved kinematics
from spectra obtained at the Keck Telescopes.
\mbh~is determined based on the single-epoch method with broad H$\beta$
emission-line width measured from Keck spectra.
Surface photometry is performed using state of the art methods,
providing a structural decomposition of the AGN host galaxies
into spheroid, disk and bar (when present), with the
spheroid component conservatively being classified as either classical
or pseudo-bulge.
Scaling relations between \mbh-and host galaxy properties
--- spatially-resolved stellar-velocity dispersion, dynamical spheroid
mass, stellar spheroid mass and spheroid luminosity --- are presented,
in comparison with quiescent galaxies and RM AGNs taken from the
literature. Our findings can be summarized as follows:

\begin{enumerate}

\item The majority of AGNs (63/66) are hosted by galaxies classified
as spiral or S0 with a high fraction of bars (26/63) and pseudo-bulges
(22/63), typical for Seyfert-type galaxies. The wide variety of
host-galaxy morphologies makes our sample ideally suited to study the
dependency of the \mbh-host-galaxy
scaling relations with other parameters such as (pseudo) bulges and
bars.

\item Tight correlations are found between \mbh~and 
spatially-resolved stellar-velocity dispersion, dynamical spheroid
mass, stellar spheroid mass and spheroid luminosity,
without significant differences in the scatter. This is contrary to
the widely accepted paradigm that the \mbh-\s~relation is the most
fundamental of all scaling relations. 

\item The intrinsic scatter of 0.2-0.4 dex
is smaller than or comparable to that of quiescent 
galaxies, showing that spiral galaxies hosting AGNs follow the same
tight \mbh~scaling relations, contrary to many literature studies.

\item We do not find any particular outliers: objects with bars,
pseudo-bulges or signs of merger activity all fall within the intrinsic
scatter of the relations. Our results rule out hierarchical assembly as the sole
origin of the \mbh-host-galaxy scaling relations and highlight the
importance of secular evolution for growing both \mbh~and spheroid.

\item Within the uncertainties, the \mbh-\s~relation of our AGNs  is indistinguishable from the relation of AGNs
with \mbh~obtained through reverberation mapping. This
indirectly validates single-epoch virial estimators of \mbh~and is consistent with
no significant selection bias for RM AGNs.

\end{enumerate}

Our results show that all the tight  correlations can be
simultaneously satisfied by AGN hosts in the 10$^7$-10$^9$
\msun~regime if  data of sufficient quality are in hand and great care is
taken when deriving host-galaxy properties. A simple explanation of the difference between our uniformly tight relations and the larger scatter found in the literature is that \s~is generally measured more accurately than the other host galaxy parameters.
The sample presented in this paper is meant to serve as a local reference point for studies of the cosmic
evolution of the correlations between host galaxy properties and
\mbh.

\begin{deluxetable*}{llccc}
\tabletypesize{\footnotesize}
\tablecolumns{5}
  \tablecaption{Fits to the Local Scaling Relations.}
\label{table:fits}
\tablehead{
  \colhead{$X$ in relation} & \colhead{Sample} & \colhead{$\alpha$} & \colhead{$\beta$} & \colhead{Scatter}\\
(1) & (2) & (3)  & (4) & (5)}
  \startdata
$\sigma /200 {\rm km\,s^{-1}}$  & AGNs (50) \& Quiescent galaxies (51) & 8.52$\pm$0.56 & 4.00$\pm$0.25   & 0.35$\pm$0.04\\
$\sigma /200 {\rm km\,s^{-1}}$  & Quiescent galaxies (51) & 8.53$\pm$0.68 & 4.55$\pm$0.29  &
                                                                0.33$\pm$0.04
                                                        \\
$\sigma /200 {\rm km\,s^{-1}}$  & AGNs (50) & 8.51$\pm$0.08 &
  4.55 (fixed quiescent)  & 0.42$\pm$0.08  \\
\hline
$\sigma /200 {\rm km\,s^{-1}}$  & AGNs (50) \& RM AGNs (29) \& Quiescent galaxies (51) & 8.50$\pm$0.46 &  4.01$\pm$0.21  & 0.37$\pm$0.03  \\
$\sigma /200 {\rm km\,s^{-1}}$  & RM AGNs (29) \& Quiescent galaxies (51) & 8.54$\pm$0.51 & 4.34$\pm$0.23  & 0.38$\pm$0.04  \\
$\sigma /200 {\rm km\,s^{-1}}$  & RM AGNs (29) & 8.45$\pm$1.15 & 3.89$\pm$0.53  & 0.42$\pm$0.06 \\
$\sigma /200 {\rm km\,s^{-1}}$  & RM AGNs (29) & 8.57$\pm$0.09 & 4.55 (fixed quiescent)  & 0.45$\pm$0.06  \\
\hline
\hline
$M_{\rm sph}/10^{11} M_{\odot}$  & Quiescent galaxies (52)  & 
    8.78$\pm$1.06 & 1.05$\pm$0.1 &
                                                            0.43$\pm$0.05
                                                            \\                                                            
\hline
                                                            $M_{\rm sph, dyn}/10^{11} M_{\odot}$  & AGNs (50) \& Quiescent galaxies (52) & 8.76$\pm$0.69 & 0.97$\pm$0.06 &
                                                            0.37$\pm$0.04
                                                \\
$M_{\rm sph, dyn}/10^{11} M_{\odot}$  & AGNs (50) & 8.77$\pm$0.07 &
1.05 (fixed quiescent) &
                                                            0.14$\pm$0.1
                                                            \\                                                            
  \hline
$M_{\rm sph, stellar}/10^{11} M_{\odot}$  & AGNs (63) \& Quiescent galaxies (52) & 8.72$\pm$0.7 & 0.97$\pm$0.07 &
                                                            0.39$\pm$0.04
                                                 \\
$M_{\rm sph+bar,  stellar}/10^{11} M_{\odot}$  & AGNs (sph+bar) (63) \& Quiescent galaxies
                  (52) & 8.71$\pm$0.74 & 1.05$\pm$0.07 &
                                                            0.38$\pm$0.04
                                                 \\
$M_{\rm sph, stellar}/10^{11} M_{\odot}$  & AGNs (63) & 8.59$\pm$0.06
& 1.05
                                                            (fixed
                                                            quiescent) &
                                                            0.12$\pm$0.1
                                                 \\
\hline
\hline
$L_{\rm sph, I} /10^{11} L_{\odot}$  & AGNs (66) \& Quiescent
                                         galaxies (51) &   9.06$\pm$0.73  & 1.03$\pm$0.07   & 0.39$\pm$0.04  \\
$L_{\rm sph, I}/10^{11} L_{\odot}$  & AGNs (sph+bar) (66) \& Quiescent
                                      galaxies (51) &   9.05$\pm$0.76
                             & 1.11$\pm$0.07   & 0.40$\pm$0.04  \\
$L_{\rm sph, I}/10^{11} L_{\odot}$  & Quiescent galaxies (51) &   9.11$\pm$1.15  & 1.02$\pm$0.11   & 0.47$\pm$0.06  \\
$L_{\rm sph, I}/10^{11} L_{\odot}$  & AGNs (66) &   8.88$\pm$0.05 &
1.02 (fixed
                                                       quiescent)
                                              & 0.08$\pm$0.07  \\
                                              \enddata
                                              \tablecomments{
For consistency, all fits were calculated as part of this paper,
  including those to RM AGNs and quiescent galaxies alone.                                       
                                             The relations plotted as dashed lines in Fig.~\ref{figure:mbh}
                                             correspond to the ones given in sample ``AGN \& Quiescent galaxies.''
(Note that the small scatter in the \mbh-$L_{\rm sph, I}$
  relation implies that the scatter is dominated by the measurement uncertainties.)                                            
Col. (1): Scaling relation of the form $\log (M_{\rm BH}/M_{\odot}) =
\alpha + \beta \log X$ with $X$ given in the table. 
Col. (2): Sample used for fitting  (AGNs = the AGNs in this paper;
quiescent = elliptical galaxies and classical bulges from
\citet{kor13};
RM AGNs = AGNs with \mbh~determined from reverberation-mapping taken
from \citet{woo15}.
In parantheses, number of galaxies  in each sample
are given.
Col. (3): Mean and uncertainty of the best fit intercept.
Col. (4): Mean and uncertainty of the best fit slope.
Col. (5): Mean and uncertainty of the best fit intrinsic scatter.
}
\end{deluxetable*}
    
\section*{Acknowledgements}
We thank the anonymous referee for their thorough report
  and many valuable comments that helped to improve the paper.
We thank Stephane Courteau, Alessandra Lamastra, Michael McDonald, Nicola Menci,
Anowar Shajib, Daeseong Park and Jong-Hak Woo for helpful
discussions.
VNB, IS and TS gratefully acknowledge
assistance from National Science Foundation (NSF) Research at
Undergraduate Institutions (RUI) grants AST-1312296 and AST-1909297.
Note that findings and conclusions do not necessarily represent views
of the NSF.
TT acknowledges support by NSF through grant AST-1907208, and by the
Packard Foundation through a Packard Research Fellowship.
IS is supported by the German Research Foundation (DFG, German Research Foundation) under Germany’s Excellence Strategy - EXC 2121 “Quantum Universe”- 390833306.
Based on observations obtained with the Hubble Space Telescope and
supported by a Space Telescope Science Institute (STScI) grant
associated with program HST-GO-15215.
Support for Program number HST-GO-15215 was provided by NASA
through a grant from the Space Telescope Science Institute, which is
operated by the Association of Universities for Research in Astronomy,
Incorporated, under NASA contract NAS5-26555.
VNB also gratefully acknowledges support by NASA grant 80NSSC19K1016.
Based on observations obtained at the international Gemini Observatory, a program of NSF’s NOIRLab
(processed using the Gemini IRAF package), which is managed by the
Association of Universities for Research in Astronomy (AURA) under a
cooperative agreement with the National Science Foundation on behalf
of the Gemini Observatory partnership: the National Science Foundation
(United States), National Research Council (Canada), Agencia Nacional
de Investigaci\'{o}n y Desarrollo (Chile), Ministerio de Ciencia,
Tecnolog\'{i}a e Innovaci\'{o}n (Argentina), Minist\'{e}rio da
Ci\^{e}ncia, Tecnologia, Inova\c{c}\~{o}es e Comunica\c{c}\~{o}es
(Brazil), and Korea Astronomy and Space Science Institute (Republic of
Korea).
Based on observations obtained
at the W. M. Keck Observatory, which is operated as a scientific
partnership among Caltech, the University of California, and the
National Aeronautics and Space Administration (NASA). The Observatory
was made possible by the generous financial support of
the W. M. Keck Foundation.
The authors recognize and acknowledge the very significant cultural
role and reverence that the summit of Mauna Kea has always had within
the indigenous Hawaiian community.  We are most fortunate to have the
opportunity to conduct observations from this mountain.
This research has made use of the Dirac computer cluster at the
California Polytechnic State University in San Luis Obispo,
maintained by Dr. Brian Granger and Dr. Ashley Ringer McDonald,
and the Hoffman2 Cluster at the University of California Los Angeles, managed and operated by the IDRE Research Technology Group under the direction of Lisa Snyder.
This research
has made use of the public archive of the Sloan Digital Sky Survey
(SDSS) and the NASA/IPAC Extragalactic Database (NED) which is
operated by the Jet Propulsion Laboratory, California Institute of
Technology, under contract with the National Aeronautics and Space
Administration.
This research has made use of the NASA/IPAC Infrared Science Archive, which is funded by the National Aeronautics and Space Administration and operated by the California Institute of Technology.
We thank Gemini staff observers K. Chibocas, W. Fraser, L. Fuhrmann, T. Geballe, M. Hoenig, J. Miller,
S. Pakzad, R. Pike, M. Schwamb, O. Smirnova, and A. Smith for obtaining these data in queue mode.

\facilities{HST (WFC3), Keck:I (LRIS), Gemini:Gillett (NIRI), Sloan}

\software{
  astropy \citep{ast13}
  EMCEE \citep{for13}
  GALFIT \citep{pen02},
  IRAF
  \citep{tod86,tod93},
  L.A. Cosmic \citep{vanDok01},
  lenstronomy \citep{bir18},
  Matplotlib \citep{hun07}
  Particle Swarm Optimizer \citep{ken01},
  photutils \citep{bra16},
  PyRAF (Science Software Branch at STScI 2012),
    PySynphot \citep{lim15},
    Python libraries \citep{van09},
    SciPy \citep{vir20}
}

\appendix

\begin{longrotatetable}
\begin{deluxetable*}{l|cccc|cccc|cccccccccc}
\tabletypesize{\footnotesize}
\tablecolumns{19}
  \tablecaption{Surface-Photometry Fitting Results.}
  \label{fittingresults}
\tablehead{
 & \multicolumn{4}{c|}{HST $I$-band} &
  \multicolumn{4}{c|}{Gemini $Ks$-band} & \multicolumn{10}{c}{HST} \\
Object & AGN & Spheroid & Disk
& Bar & AGN & Spheroid & Disk
& Bar  & $n_{\rm sph}$ &
$R_{\rm sph}$ & $PA_{\rm sph}$ & $q_{\rm sph}$ &
$R_{\rm disk}$ & $PA_{\rm disk}$ & $q_{\rm disk}$ &
$R_{\rm bar}$ & $PA_{\rm bar}$ & $q_{\rm bar}$ \\
& (mag) & (mag) & (mag) & (mag) 
& (mag) & (mag) & (mag) & (mag) 
&  & ($\prime\prime$) & ($\deg$) & &
($\prime\prime$) & ($\deg$) & &
($\prime\prime$) & ($\deg$) \\
(1) & (2) & (3)  & (4) & (5) & (6) & (7) & (8) & (9)  & (10) & (11) &
(12) & (13)  & (14) & (15) & (16) & (17) & (18) & (19)}
\startdata
0013-0951 & 18.3 & 18.0 & 15.9 & ... & 16.0 & 17.5 & 15.0 & ... & 1.2 & 0.36 & 12.6 & 0.77 & 5.19 & 39.2 & 0.52 & ... & ... & ... \\ 
0038+0034 & 17.9 & 16.3 & 18.3 & ... & ... & ... & ... & ... & 4.5 & 2.39 & 158.0 & 0.82 & 2.39 & 104.9 & 0.44 & ... & ... & ... \\ 
0109+0059 & 18.9 & 18.3 & 16.9 & 18.2 & 17.8 & 17.1 & 15.9 & 17.2 & 1.6 & 0.2 & 74.8 & 0.63 & 3.2 & 80.2 & 0.53 & 1.11 & 78.2 & 0.31 \\ 
0121-0102 & 16.7 & 18.4 & 14.8 & 16.6 & 15.1 & 17.0 & 14.1 & 15.6 & 1.0 & 0.42 & 33.6 & 0.95 & 5.35 & 67.1 & 0.95 & 2.71 & 66.6 & 0.27 \\ 
0150+0057 & 19.1 & 17.6 & 15.5 & 17.2 & 17.0 & 16.8 & 14.6 & 16.2 & 1.0 & 0.39 & 63.9 & 0.84 & 3.92 & 54.8 & 0.8 & 1.3 & 172.0 & 0.4 \\ 
0206-0017 & 19.0 & 14.1 & 14.2 & ... & 16.0 & 13.1 & 13.5 & ... & 3.8 & 3.27 & 1.7 & 0.72 & 9.42 & 174.6 & 0.56 & ... & ... & ... \\ 
0212+1406 & 18.4 & 16.8 & 15.6 & 18.4 & ... & ... & ... & ... & 1.0 & 0.75 & 0.3 & 0.64 & 4.22 & 38.3 & 0.42 & 0.37 & 58.9 & 0.39 \\ 
0301+0110 & 18.1 & 16.9 & 17.7 & ... & ... & ... & ... & ... & 4.0 & 1.2 & 162.0 & 0.65 & 3.35 & 77.1 & 0.65 & ... & ... & ... \\ 
0301+0115 & 18.0 & 18.1 & 16.8 & 18.1 & 15.9 & 16.4 & 15.7 & 17.3 & 1.0 & 0.23 & 88.3 & 0.89 & 2.95 & 118.5 & 0.77 & 1.46 & 145.0 & 0.34 \\ 
0336-0706 & 20.7 & 17.2 & 16.4 & ... & ... & ... & ... & ... & 1.0 & 0.77 & 15.6 & 0.67 & 5.4 & 4.1 & 0.24 & ... & ... & ... \\ 
0353-0623 & 18.4 & 17.3 & 16.9 & 18.4 & ... & ... & ... & ... & 1.0 & 0.94 & 159.3 & 0.73 & 4.75 & 177.1 & 0.34 & 0.27 & 179.7 & 0.34 \\ 
0737+4244 & 19.7 & 17.5 & 16.6 & ... & ... & ... & ... & ... & 2.7 & 0.11 & 158.3 & 0.8 & 2.85 & 19.2 & 0.56 & ... & ... & ... \\ 
0802+3104 & 17.2 & 17.4 & 15.6 & 17.7 & ... & ... & ... & ... & 1.1 & 0.28 & 66.1 & 0.85 & 3.27 & 74.0 & 0.84 & 1.01 & 140.0 & 0.4 \\ 
0811+1739 & 19.1 & 17.7 & 16.2 & 17.9 & ... & ... & ... & ... & 1.4 & 0.38 & 105.9 & 0.79 & 4.48 & 97.3 & 0.73 & 2.88 & 71.4 & 0.39 \\ 
0813+4608 & 20.1 & 16.6 & 16.6 & 17.0 & 17.8 & 15.7 & 15.7 & 16.1 & 2.6 & 0.67 & 105.8 & 0.89 & 5.14 & 106.3 & 0.69 & 3.58 & 118.4 & 0.37 \\ 
0845+3409 & 20.1 & 16.9 & 15.9 & 18.1 & 17.4 & 15.7 & 15.2 & 17.2 & 3.9 & 1.09 & 16.1 & 0.89 & 7.0 & 77.1 & 0.89 & 1.66 & 10.0 & 0.27 \\ 
0857+0528 & 18.0 & 17.7 & 15.6 & ... & ... & ... & ... & ... & 1.2 & 0.38 & 131.1 & 0.71 & 3.56 & 133.6 & 0.59 & ... & ... & ... \\ 
0904+5536 & 16.7 & 16.4 & 16.6 & ... & ... & ... & ... & ... & 1.8 & 1.35 & 84.7 & 0.57 & 7.42 & 46.2 & 0.49 & ... & ... & ... \\ 
0909+1330 & 20.6 & 17.7 & 15.4 & 17.3 & ... & ... & ... & ... & 1.5 & 0.64 & 38.0 & 0.88 & 7.61 & 45.7 & 0.75 & 4.61 & 70.1 & 0.24 \\ 
0921+1017 & 18.7 & 15.7 & 14.7 & ... & ... & ... & ... & ... & 5.0 & 3.88 & 85.6 & 0.97 & 3.97 & 108.1 & 0.97 & ... & ... & ... \\ 
0923+2254 & 16.4 & 15.5 & 14.2 & 16.4 & ... & ... & ... & ... & 1.4 & 0.95 & 173.0 & 0.79 & 10.9 & 176.3 & 0.56 & 6.4 & 160.7 & 0.25 \\ 
0923+2946 & 21.1 & 15.5 & ... & ... & ... & ... & ... & ... & 4.8 & 2.37 & 111.9 & 0.91 & ... & ... & ... & ... & ... & ... \\ 
0927+2301 & 17.7 & 14.8 & 13.5 & ... & ... & ... & ... & ... & 1.4 & 1.24 & 99.0 & 0.71 & 7.46 & 66.4 & 0.71 & ... & ... & ... \\ 
0932+0233 & 18.3 & 16.9 & 16.2 & ... & ... & ... & ... & ... & 1.1 & 0.68 & 140.1 & 0.7 & 4.37 & 145.9 & 0.7 & ... & ... & ... \\ 
0936+1014 & 16.9 & 18.0 & 15.2 & ... & ... & ... & ... & ... & 1.4 & 0.65 & 19.3 & 0.33 & 6.15 & 23.0 & 0.31 & ... & ... & ... \\ 
1029+1408 & 18.3 & 15.4 & 16.6 & ... & ... & ... & ... & ... & 4.3 & 2.42 & 11.8 & 0.48 & 5.87 & 3.2 & 0.44 & ... & ... & ... \\ 
1029+2728 & 19.8 & 16.2 & 16.5 & ... & ... & ... & ... & ... & 3.1 & 0.78 & 175.7 & 0.88 & 3.21 & 12.2 & 0.84 & ... & ... & ... \\ 
1029+4019 & 17.4 & 17.7 & 16.2 & ... & ... & ... & ... & ... & 1.1 & 0.42 & 109.2 & 0.64 & 2.5 & 84.1 & 0.64 & ... & ... & ... \\ 
1042+0414 & 18.8 & 17.6 & 16.5 & 18.1 & ... & ... & ... & ... & 1.6 & 0.25 & 114.0 & 0.85 & 3.21 & 117.6 & 0.79 & 2.02 & 134.8 & 0.38 \\ 
1043+1105 & 16.9 & 16.8 & ... & ... & ... & ... & ... & ... & 3.1 & 2.3 & 121.3 & 0.9 & ... & ... & ... & ... & ... & ... \\ 
1058+5259 & 18.3 & 16.9 & 16.7 & 17.3 & ... & ... & ... & ... & 1.9 & 0.63 & 22.6 & 0.86 & 5.94 & 32.4 & 0.73 & 2.58 & 44.3 & 0.33 \\ 
1101+1102 & 19.1 & 15.4 & 15.7 & ... & ... & ... & ... & ... & 5.0 & 3.0 & 173.8 & 0.77 & 4.59 & 144.1 & 0.48 & ... & ... & ... \\ 
1104+4334 & 20.6 & 15.8 & 18.7 & 18.6 & ... & ... & ... & ... & 4.5 & 2.79 & 21.2 & 0.81 & 2.79 & 70.3 & 0.49 & 1.46 & 32.9 & 0.31 \\ 
1137+4826 & 22.9 & 17.5 & 17.5 & ... & ... & ... & ... & ... & 2.6 & 0.24 & 83.0 & 0.76 & 1.29 & 103.5 & 0.7 & ... & ... & ... \\ 
1143+5941 & 18.7 & 17.5 & 16.5 & 17.5 & ... & ... & ... & ... & 2.4 & 0.5 & 118.8 & 0.9 & 6.72 & 7.0 & 0.65 & 4.26 & 5.9 & 0.3 \\ 
1144+3653 & 16.3 & 15.8 & 15.0 & ... & ... & ... & ... & ... & 1.4 & 1.2 & 29.2 & 0.79 & 7.87 & 11.5 & 0.79 & ... & ... & ... \\ 
1145+5547 & 19.3 & 18.2 & 15.3 & 18.0 & ... & ... & ... & ... & 1.0 & 0.52 & 54.3 & 0.68 & 7.0 & 67.8 & 0.66 & 2.1 & 58.1 & 0.33 \\ 
1147+0902 & 16.6 & 15.9 & 18.2 & ... & ... & ... & ... & ... & 3.8 & 2.01 & 105.6 & 0.72 & 2.01 & 138.2 & 0.36 & ... & ... & ... \\ 
1205+4959 & 17.2 & 16.6 & 15.7 & ... & ... & ... & ... & ... & 2.0 & 0.68 & 169.9 & 0.88 & 4.41 & 162.1 & 0.88 & ... & ... & ... \\ 
1206+4244 & 17.4 & 17.2 & 15.5 & 17.1 & ... & ... & ... & ... & 1.0 & 0.64 & 138.0 & 0.98 & 7.0 & 104.9 & 0.98 & 2.57 & 134.7 & 0.33 \\ 
1216+5049 & 18.4 & 15.0 & 14.9 & ... & ... & ... & ... & ... & 3.3 & 2.75 & 79.2 & 0.5 & 8.03 & 72.9 & 0.33 & ... & ... & ... \\ 
1223+0240 & 16.6 & 15.3 & 15.0 & ... & ... & ... & ... & ... & 5.0 & 3.51 & 167.4 & 0.83 & 4.3 & 174.2 & 0.77 & ... & ... & ... \\ 
1246+5134 & 19.1 & 17.9 & 17.1 & ... & ... & ... & ... & ... & 2.0 & 0.4 & 87.0 & 0.52 & 2.88 & 91.7 & 0.25 & ... & ... & ... \\ 
1306+4552 & 20.9 & 18.3 & 15.6 & 17.1 & ... & ... & ... & ... & 1.0 & 0.3 & 171.6 & 0.91 & 5.05 & 127.4 & 0.91 & 2.88 & 152.9 & 0.36 \\ 
1307+0952 & 19.1 & 17.4 & 15.3 & 19.4 & ... & ... & ... & ... & 1.0 & 0.56 & 136.1 & 0.85 & 4.59 & 174.0 & 0.66 & 0.27 & 86.6 & 0.4 \\ 
1312+2628 & 17.5 & 17.9 & 15.5 & 17.9 & ... & ... & ... & ... & 1.0 & 0.45 & 173.8 & 0.88 & 6.03 & 169.5 & 0.88 & 2.7 & 0.0 & 0.22 \\ 
1405-0259 & 18.8 & 17.4 & 15.6 & ... & ... & ... & ... & ... & 1.0 & 0.77 & 59.3 & 0.59 & 5.01 & 64.6 & 0.4 & ... & ... & ... \\ 
1416+0137 & 18.5 & 15.9 & 15.2 & ... & ... & ... & ... & ... & 2.4 & 1.57 & 158.7 & 0.85 & 6.75 & 132.4 & 0.67 & ... & ... & ... \\ 
1419+0754 & 17.6 & 16.1 & 14.5 & ... & ... & ... & ... & ... & 1.9 & 0.98 & 21.8 & 0.72 & 6.45 & 18.5 & 0.68 & ... & ... & ... \\ 
1434+4839 & 17.7 & 16.3 & 15.0 & 16.4 & ... & ... & ... & ... & 1.5 & 0.88 & 153.2 & 0.89 & 6.29 & 156.4 & 0.83 & 3.98 & 150.5 & 0.32 \\ 
1545+1709 & 18.7 & 16.2 & 16.9 & ... & ... & ... & ... & ... & 5.0 & 1.49 & 62.0 & 0.59 & 2.42 & 60.1 & 0.2 & ... & ... & ... \\ 
1557+0830 & 17.9 & 16.8 & ... & ... & ... & ... & ... & ... & 2.5 & 1.16 & 58.7 & 0.78 & ... & ... & ... & ... & ... & ... \\ 
1605+3305 & 17.9 & 16.9 & 16.5 & ... & ... & ... & ... & ... & 1.2 & 0.68 & 97.0 & 0.66 & 3.82 & 82.8 & 0.61 & ... & ... & ... \\ 
1606+3324 & 18.6 & 16.3 & 16.5 & ... & ... & ... & ... & ... & 3.3 & 1.16 & 18.0 & 0.7 & 4.29 & 20.2 & 0.48 & ... & ... & ... \\ 
1611+5211 & 18.8 & 15.8 & 16.2 & ... & ... & ... & ... & ... & 3.1 & 0.75 & 126.6 & 0.81 & 4.62 & 106.3 & 0.68 & ... & ... & ... \\ 
1636+4202 & 18.4 & 15.8 & 17.1 & 18.2 & ... & ... & ... & ... & 5.0 & 3.0 & 22.4 & 0.73 & 3.0 & 9.1 & 0.45 & 0.33 & 100.2 & 0.45 \\ 
1708+2153 & 16.6 & 16.5 & 16.3 & ... & ... & ... & ... & ... & 1.9 & 1.54 & 73.5 & 0.69 & 6.95 & 70.6 & 0.49 & ... & ... & ... \\ 
2116+1102 & 18.1 & 17.6 & 16.0 & 19.0 & ... & ... & ... & ... & 1.2 & 0.5 & 50.4 & 0.87 & 5.95 & 80.7 & 0.87 & 1.96 & 86.6 & 0.34 \\ 
2140+0025 & 16.5 & 17.3 & 16.8 & ... & 15.0 & 17.0 & 15.7 & ... & 1.0 & 0.47 & 79.3 & 0.71 & 2.24 & 88.3 & 0.71 & ... & ... & ... \\ 
2215-0036 & 17.0 & 18.1 & 16.7 & ... & ... & ... & ... & ... & 1.0 & 0.5 & 82.7 & 0.46 & 3.45 & 83.1 & 0.4 & ... & ... & ... \\ 
2221-0906 & 17.6 & 17.6 & 17.3 & ... & 16.6 & 17.6 & 15.9 & ... & 2.7 & 0.86 & 24.5 & 0.69 & 2.98 & 81.6 & 0.69 & ... & ... & ... \\ 
2222-0819 & 17.4 & 18.7 & 15.6 & 17.8 & 15.0 & 25.9 & 14.7 & 16.6 & 1.1 & 0.33 & 77.7 & 0.93 & 3.98 & 78.3 & 0.93 & 1.48 & 108.2 & 0.32 \\ 
2233+1312 & 17.9 & 17.9 & 15.9 & 17.5 & 16.2 & 16.4 & 14.8 & 16.2 & 1.0 & 0.38 & 65.9 & 0.66 & 7.71 & 57.1 & 0.53 & 1.82 & 25.6 & 0.51 \\ 
2254+0046 & 17.0 & 17.4 & 17.6 & ... & ... & ... & ... & ... & 1.1 & 0.26 & 113.0 & 0.91 & 1.91 & 79.6 & 0.9 & ... & ... & ... \\ 
2327+1524 & 17.9 & 14.7 & 14.8 & ... & 15.1 & 13.5 & 14.1 & ... & 2.4 & 1.71 & 3.9 & 0.75 & 7.86 & 6.8 & 0.75 & ... & ... & ... \\ 
2351+1552 & 17.7 & 17.0 & 16.9 & ... & 16.5 & 16.1 & 15.8 & ... & 2.0 & 0.77 & 92.5 & 0.63 & 2.69 & 104.4 & 0.49 & ... & ... & ... \\ 
\enddata
  \tablecomments{Surface-photometry fitting results using lenstronomy on HST and Gemini images.
Col. (1): Target ID used throughout the text (based on R.A. and declination). 		      	   	  
Col. (2): Point-source (AGN) magnitude in HST $I$-band (uncertainty 0.1 mag).
Col. (3): Spheroid magnitude in HST $I$-band  (uncertainty 0.1 mag).
Col. (4): Disk magnitude in HST $I$-band (if present;  uncertainty 0.1 mag).
Col. (5): Bar magnitude in HST $I$-band (if present;  uncertainty 0.1 mag).
Col. (6): Point-source (AGN) magnitude in Gemini $Ks$-band (uncertainty 0.2 mag).
Col. (7):  Spheroid magnitude in Gemini $Ks$-band (uncertainty 0.2 mag).
Col. (8):  Disk magnitude in Gemini $Ks$-band (if present;  uncertainty 0.2 mag).
Col. (9):  Bar magnitude in Gemini $Ks$-band (if present;  uncertainty 0.2 mag).
Col. (10): Spheroid S{\'e}rsic index $n$  (5\% uncertainty).
Col. (11): Spheroid radius in arcseconds (10\% uncertainty).
Col. (12): Spheroid position angle (East of North) in degree (5 $\deg$ uncertainty).
Col. (13): Spheroid axis ratio $q$ (=$b/a$) (0.05 uncertainty).
Col. (14): Disk radius in arcseconds (10\% uncertainty).
Col. (15): Disk position angle (East of North) in degree (5 $\deg$ uncertainty).
Col. (16): Disk axis ratio $q$ (=$b/a$) (0.05  uncertainty).
Col. (17): Bar radius in arcseconds (10\% uncertainty).
Col. (18): Bar position angle (East of North) in degree (5 $\deg$ uncertainty).
Col. (19): Bar axis ratio $q$ (=$b/a$) (0.05  uncertainty).
}
\end{deluxetable*}
\end{longrotatetable}

\begin{figure*}
  \includegraphics[scale=0.28]{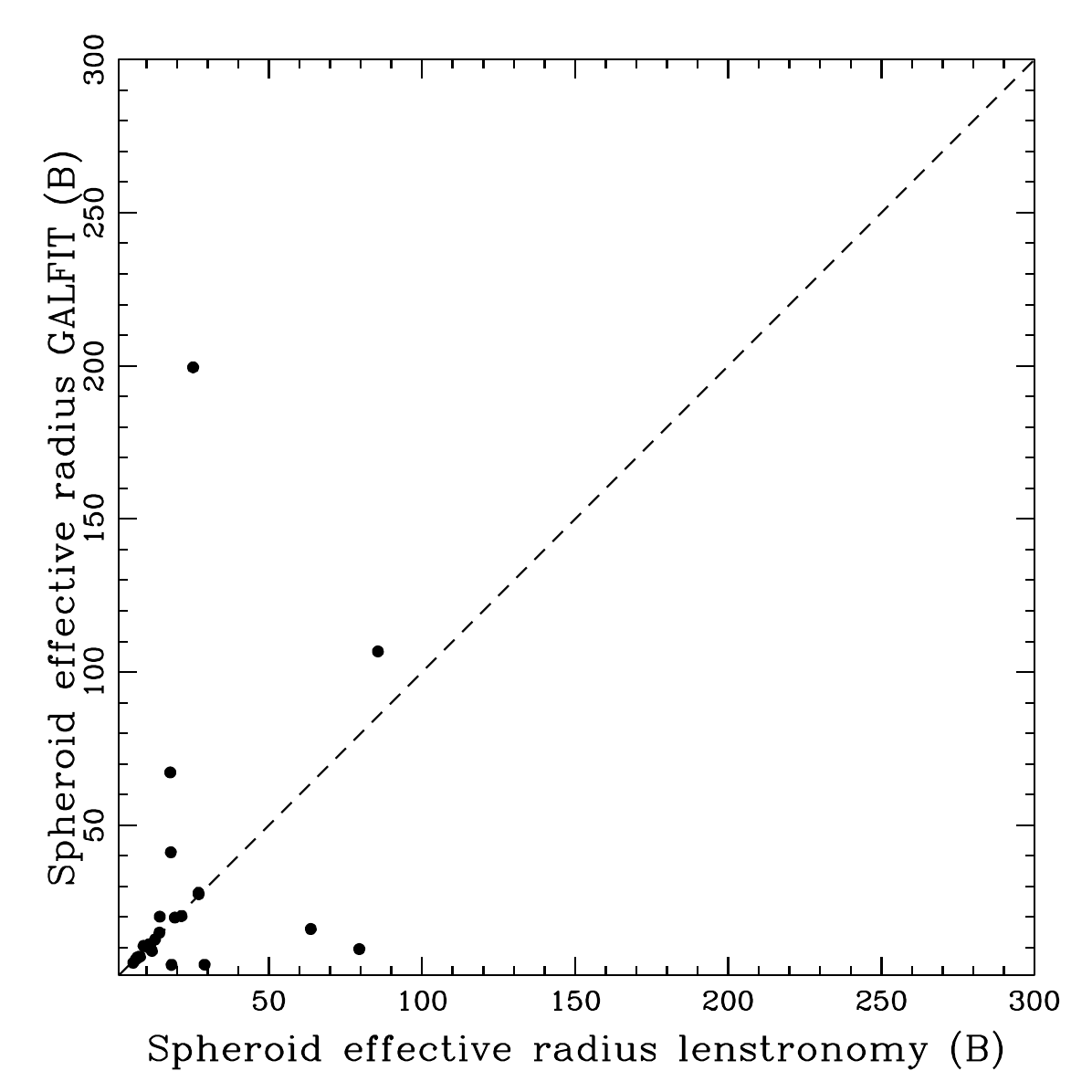}
  \includegraphics[scale=0.28]{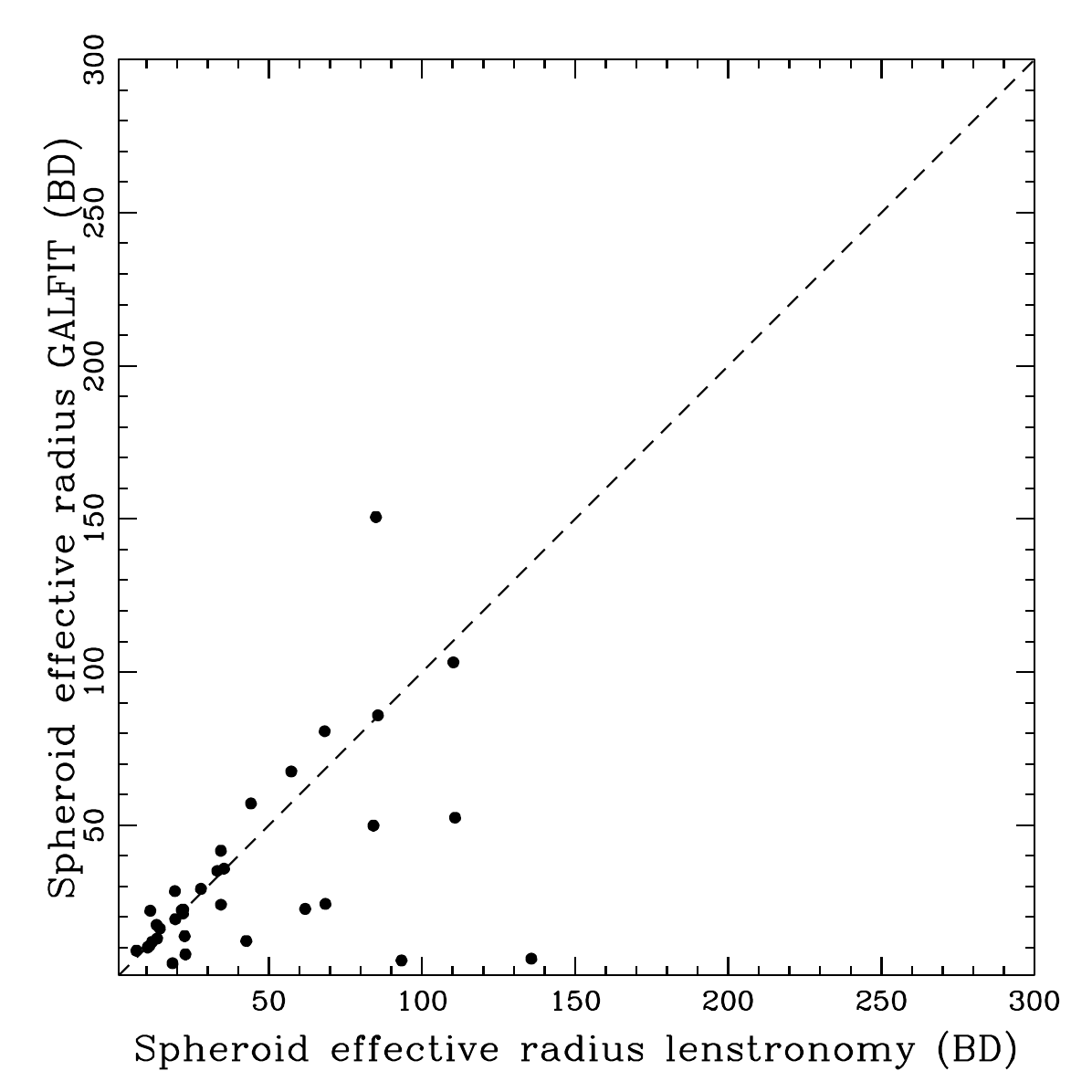}
  \includegraphics[scale=0.28]{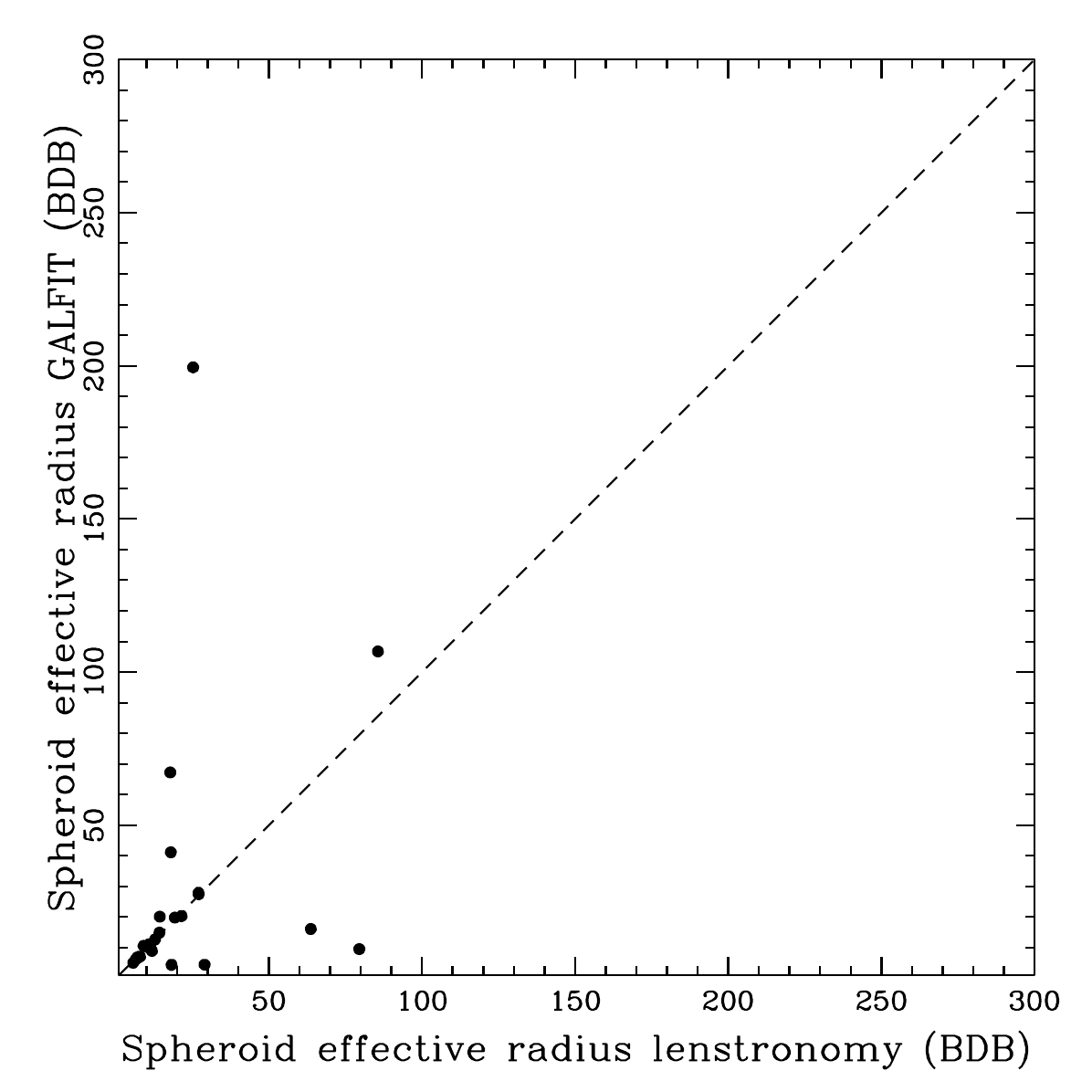}\\
  \includegraphics[scale=0.28]{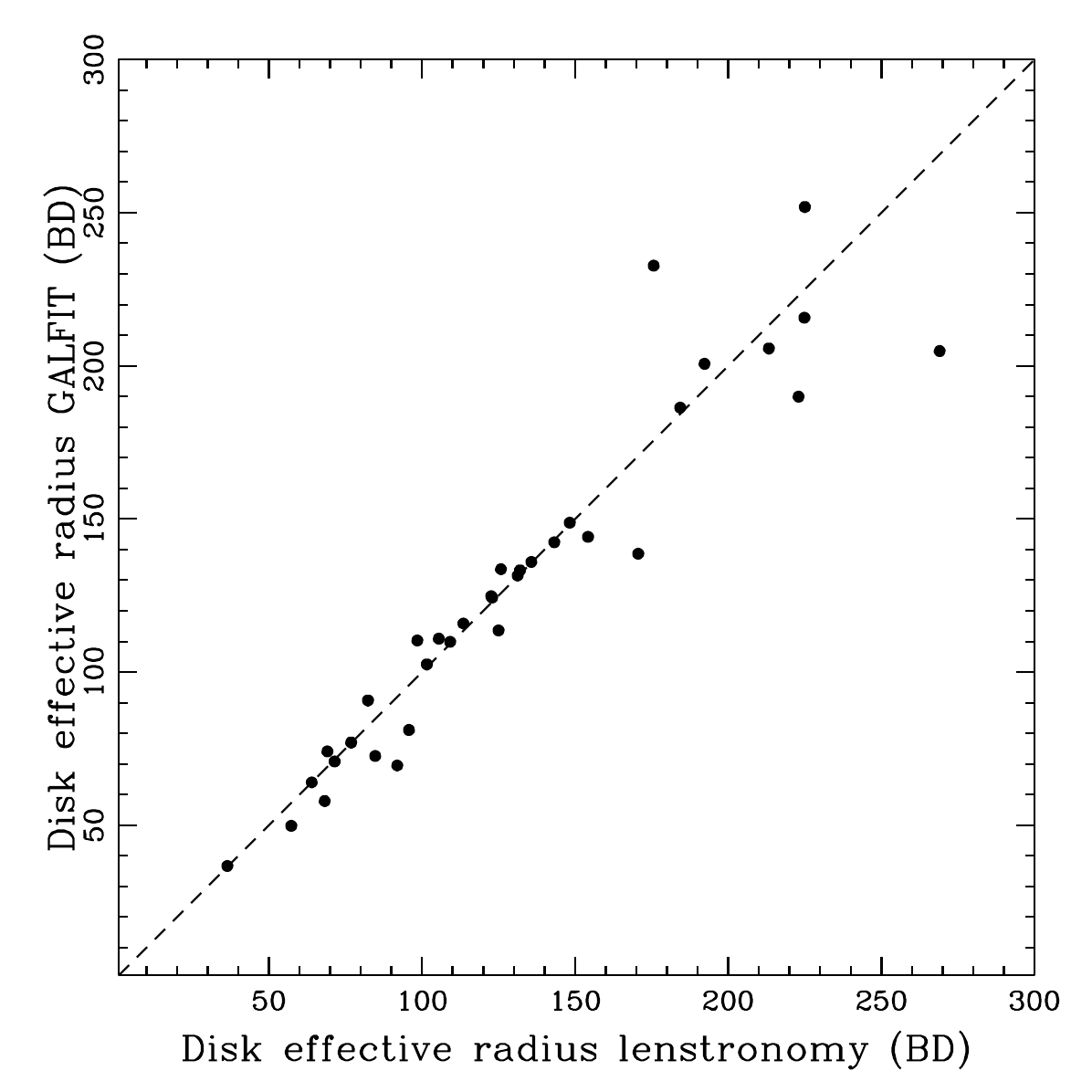}
  \includegraphics[scale=0.28]{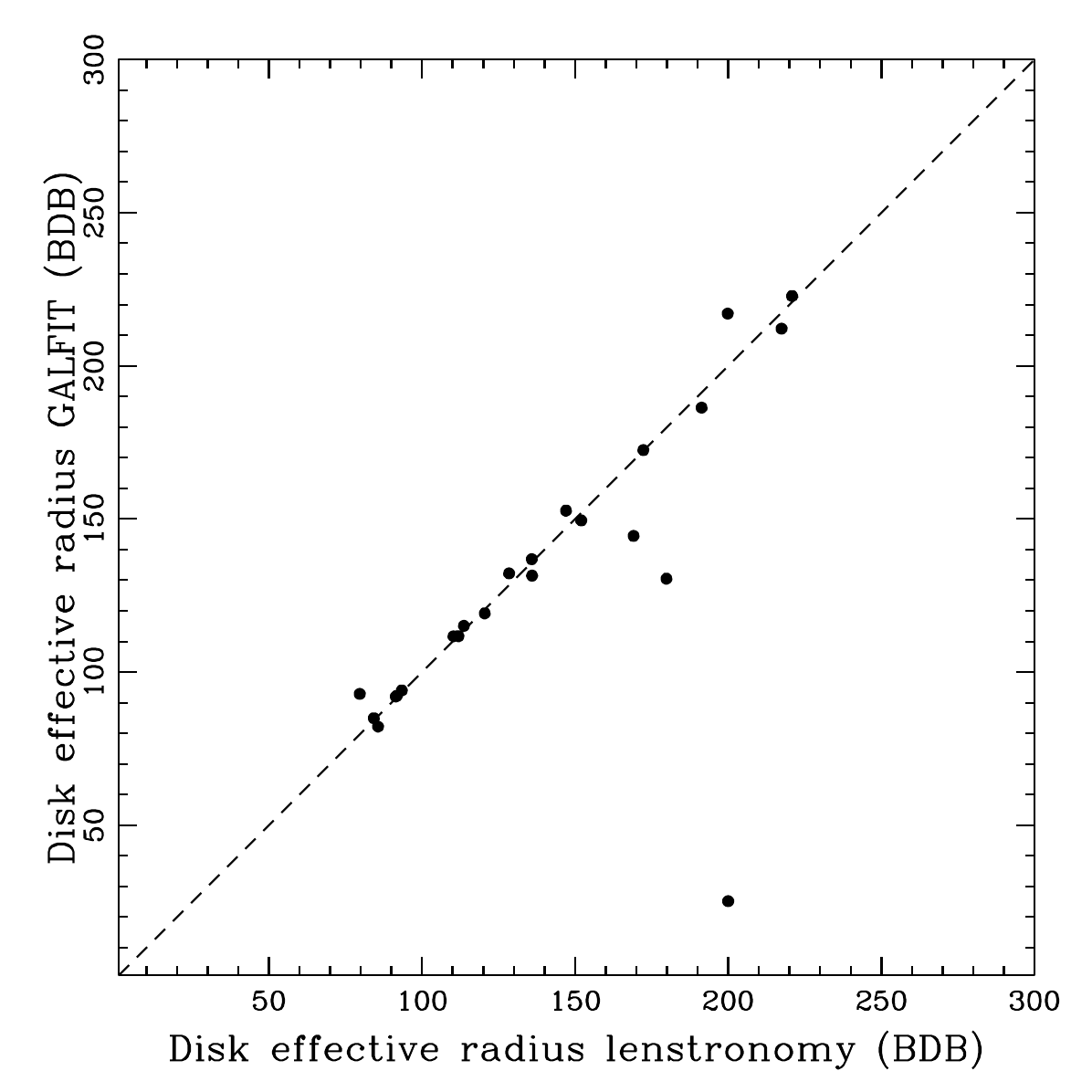}
    \includegraphics[scale=0.28]{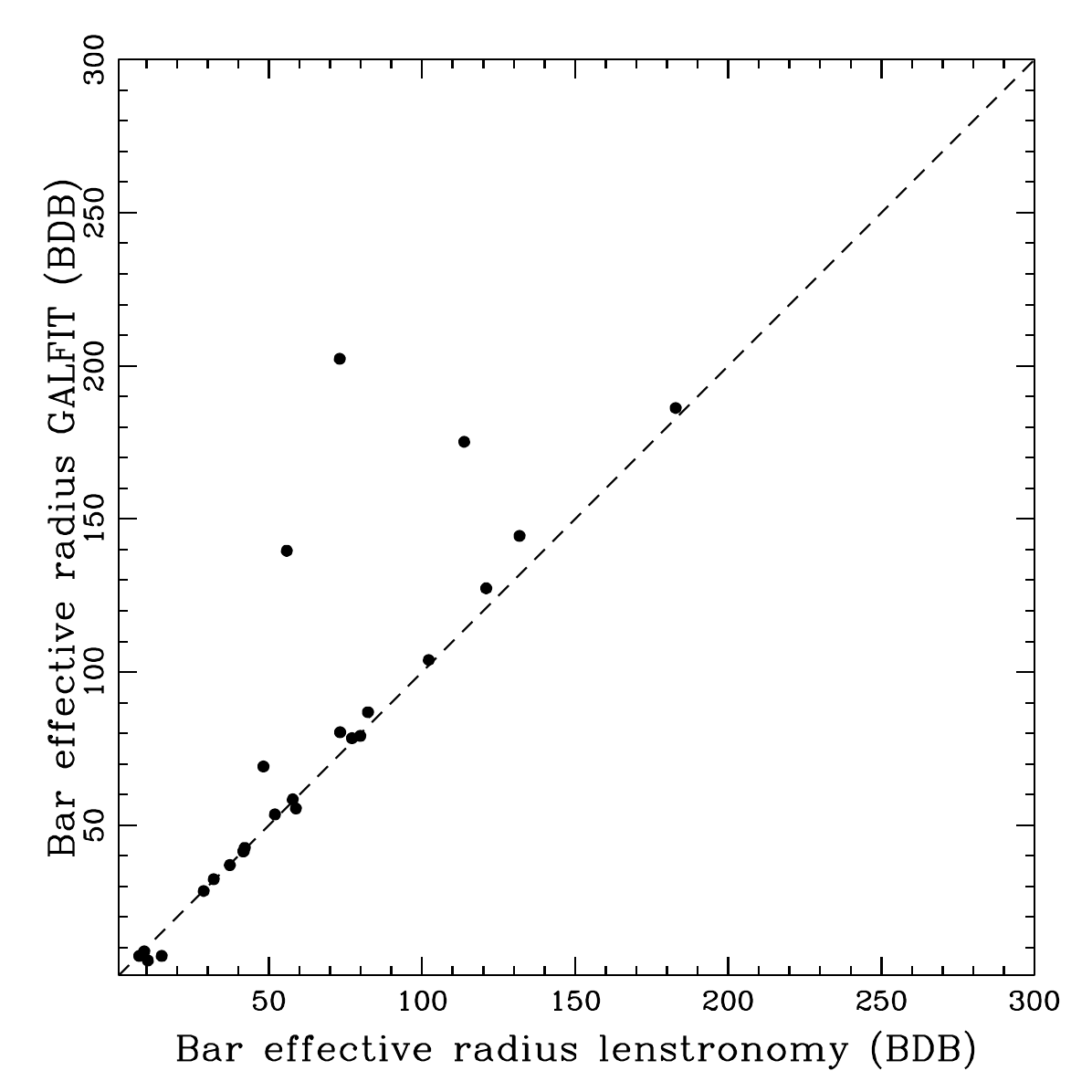}\\  
  \includegraphics[scale=0.28]{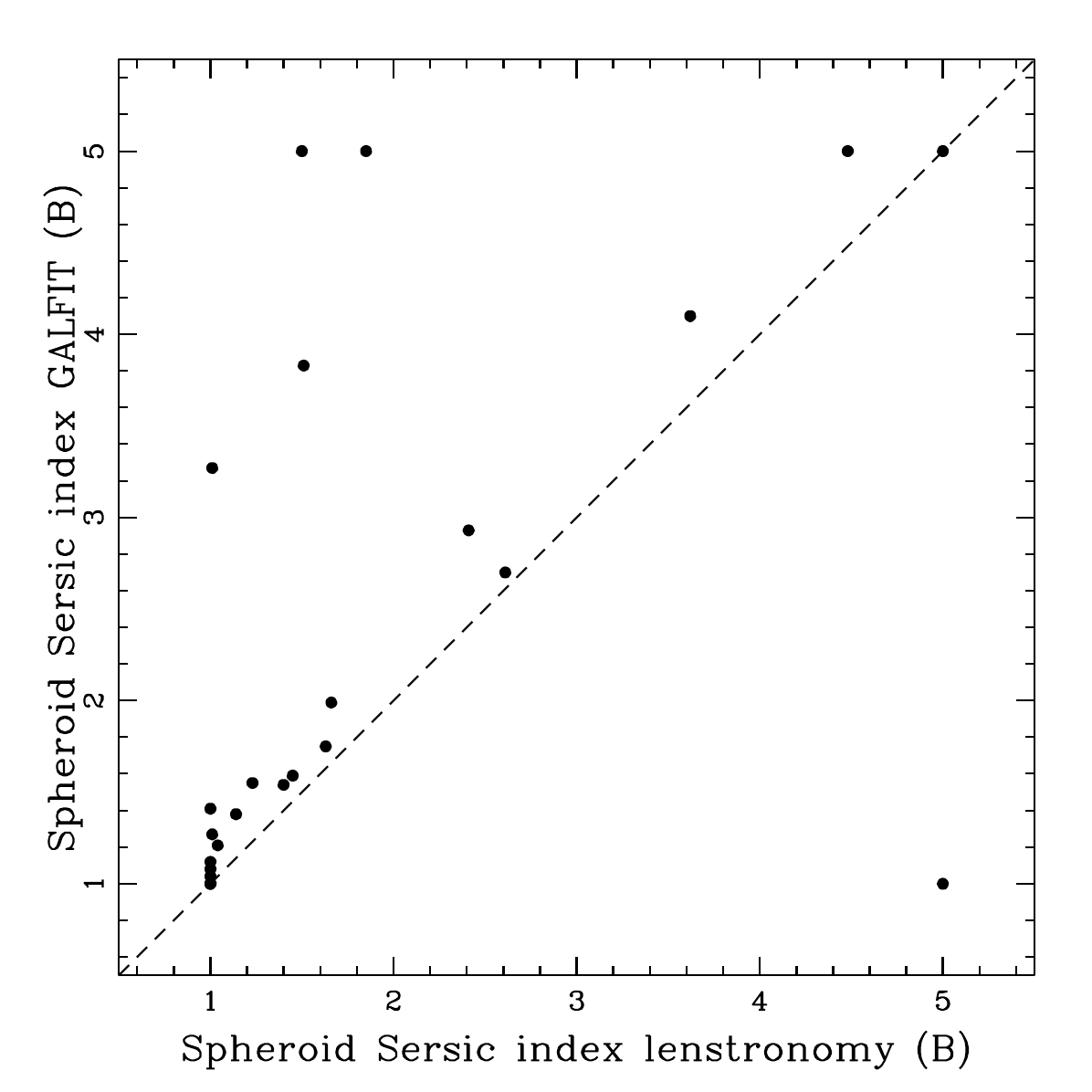}
  \includegraphics[scale=0.28]{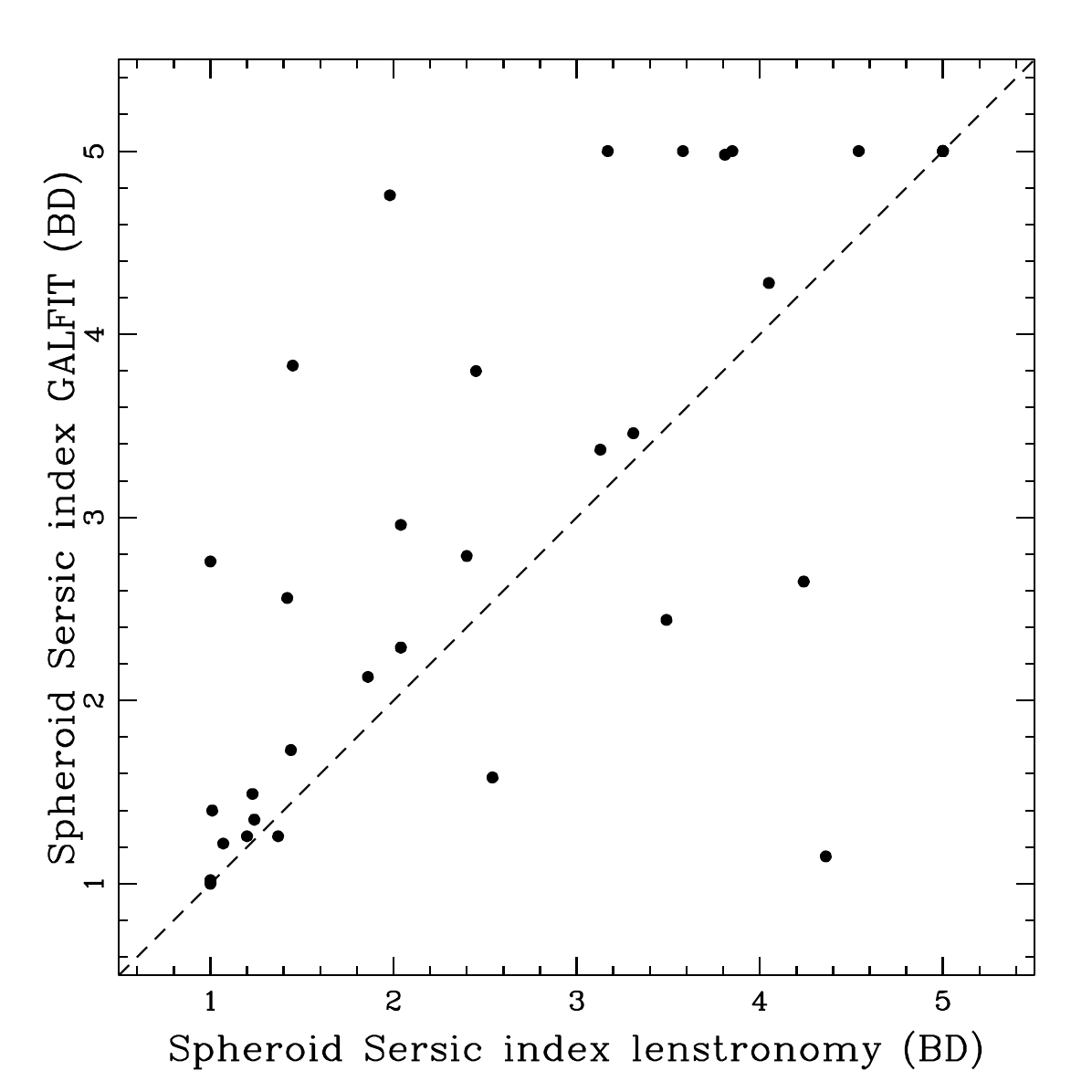}
    \includegraphics[scale=0.28]{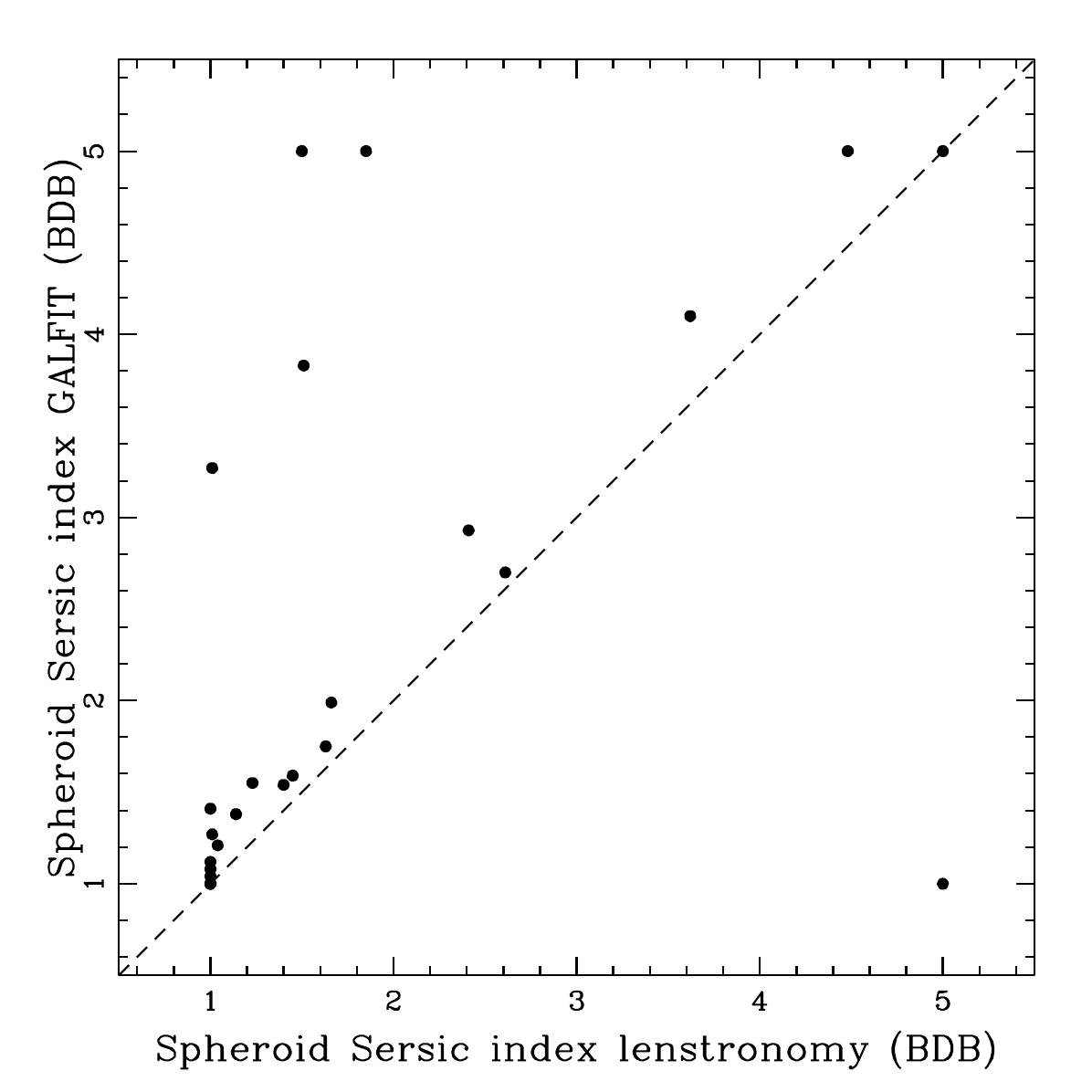}\\
    \caption{Comparison of fitting results using lenstronomy (x-axis)
      and GALFIT (y-axis) for effective radii (in pixels) and
      S{\'e}rsic index $n$ for the different components (spheroid,
      disk, bar) in the different fits. To help guide the eye, the
      dashed-line represents the 1:1 correlation.
    \label{figure:galfit}}
\end{figure*}

\end{document}